\documentclass[sigplan,screen]{acmart}

\newif\ifisdraft
\isdraftfalse

\usepackage[]{hyperref}

\usepackage{xspace}
\usepackage[T1]{fontenc}
\usepackage[utf8]{inputenc}
\usepackage{xcolor}
\usepackage{listings}

\usepackage{pgfplots}
\pgfplotsset{compat=newest}
\usetikzlibrary{decorations.pathmorphing}
\usepackage{pgfplotstable}

\usepackage{multirow}
\makeatletter
\g@addto@macro \normalsize {%
 \setlength\abovedisplayskip{1pt plus 0pt minus 0pt}%
 \setlength\belowdisplayskip{1pt plus 0pt minus 0pt}%
}
\makeatother

\usepackage{enumitem}%
\setlist[itemize]{noitemsep, topsep=2pt, leftmargin=14pt}

\newcommand{\etal}{\textit{et al.}}

\newcommand{\myparagraph}[1]{
  \noindent\textbf{\textit{#1.}}
}

\usepackage{tikz}
\newcommand*\circled[1]{\tikz[baseline=(char.base)]{
            \node[shape=circle,draw,inner sep=0.5pt] (char) {#1};}}

\renewenvironment{quote}{%
  \list{}{%
    \leftmargin0.25cm   %
    \rightmargin\leftmargin
    \topsep2pt
  }
  \item\relax
}
{\endlist}

\renewcommand{\th}[1]{{\textbf{#1}}}

\ifisdraft

\newcommand{\doremove}[2]{
    {{\colorbox{red}{\bfseries\sffamily\scriptsize\textcolor{white}{#1}}}}
    {\textcolor{lightgray}{\sf\small$\blacktriangleright~$\textit\underline{#2}$~\blacktriangleleft$}}}

\newcommand\remove[1]{\doremove{Remove}{#1}}
\else
\newcommand{\remove}[1]{}
\fi %

\definecolor{listing-comment}{rgb}{0.20,0.20,0.20}
\definecolor{listing-string}{rgb}{0, 0, 0.5}%
\definecolor{listing-keyword-traits}{rgb}{0.75, 0, 0}%
\definecolor{listing-keyword-types}{rgb}{0, 0.5, 0}%
\definecolor{listing-keyword-const}{rgb}{0, 0.5, 0}%
\definecolor{listing-keyword-macro}{rgb}{0, 0, 0.75}%
\definecolor{listing-keyword}{rgb}{0, 0, 0}%
\definecolor{listing-identifier}{rgb}{0, 0, 0.75}%

\newcommand{\lstbasicstyle}{\ttfamily\scriptsize}
\newcommand{\lstinlinebasicstyle}{\ttfamily\small}

\newcommand{\lstnumberstyle}{\tiny\color{gray}}

\newcommand{\lstkeywordstyle}{\bfseries\color{listing-keyword}}

\lstdefinestyle{common}
{
  firstnumber=1
  , basicstyle=\lstbasicstyle
  , identifierstyle=\color{listing-identifier}
  , commentstyle=\itshape\color{listing-comment}
  , stringstyle=\color{listing-string}
  , keywordstyle=\lstkeywordstyle
  , extendedchars=false
  , tabsize=2
  , showtabs=false
  , showspaces=false
  , showstringspaces=false
  , numbers=left
  , stepnumber=2
  , numberfirstline=false,
  , numberstyle=\lstnumberstyle
  , numbersep=5pt
  , captionpos=b
  , breaklines=true
  , breakatwhitespace=true
  , breakautoindent=true
  , frame=none
  , rulecolor=\color{black}
  , columns=[c]fixed
  , keepspaces=true
 , escapeinside={(*@}{@*)}
}

\lstdefinestyle{coloredProlog}
{
  language=Prolog
  , style = common
  , morecomment=[l]{\//},
}

\lstnewenvironment{codeprolog}
  {\lstset{style=coloredProlog}}
  {}

\lstnewenvironment{codeprologfloat}
  {\lstset{style=coloredProlog,float=ht, belowskip=-0.5\baselineskip}}
  {}

\lstdefinestyle{coloredC}
{
  language=C
  , style = common
  , morekeywords={uint8_t, uint16_t, uint32_t, uint64_t, inline, size_t, vaddr_t, flags_t, paddr_t}
  , morecomment=[l]{\//},
}

\lstnewenvironment{codecfloat}[1]
  {\lstset{style=coloredC,float=ht, belowskip=-0.5\baselineskip, #1}}
  {}

\lstnewenvironment{codecfloat*}[1]
  {\lstset{style=coloredC,float=*, belowskip=-0.5\baselineskip, #1}}
  {}

\lstdefinelanguage{Rust}{%
    sensitive%
  , morecomment=[l]{//}%
  , morecomment=[s]{/*}{*/}%
  , moredelim=[s][{\itshape\color[rgb]{0,0,0.75}}]{\#[}{]}%
  , morestring=[b]{"}%
  , alsodigit={}%
  , alsoother={}%
  , alsoletter={!}%
  , morekeywords={break, continue, else, for, if, in, loop, match, return, while}  %
  , morekeywords={as, const, let, move, mut, ref, static}  %
  , morekeywords={dyn, enum, fn, impl, Self, self, struct, trait, type, union, use, where}  %
  , morekeywords={crate, extern, mod, pub, super}  %
  , morekeywords={unsafe}  %
  , morekeywords={abstract, alignof, become, box, do, final, macro, offsetof, override, priv, proc, pure, sizeof, typeof, unsized, virtual, yield}  %
  , morekeywords=[2]{Add, AddAssign, Any, AsciiExt, AsInner, AsInnerMut, AsMut, AsRawFd, AsRawHandle, AsRawSocket, AsRef, Binary, BitAnd, BitAndAssign, Bitor, BitOr, BitOrAssign, BitXor, BitXorAssign, Borrow, BorrowMut, Boxed, BoxPlace, BufRead, BuildHasher, CastInto, CharExt, Clone, CoerceUnsized, CommandExt, Copy, Debug, DecodableFloat, Default, Deref, DerefMut, DirBuilderExt, DirEntryExt, Display, Div, DivAssign, DoubleEndedIterator, DoubleEndedSearcher, Drop, EnvKey, Eq, Error, ExactSizeIterator, ExitStatusExt, Extend, FileExt, FileTypeExt, Float, Fn, FnBox, FnMut, FnOnce, Freeze, From, FromInner, FromIterator, FromRawFd, FromRawHandle, FromRawSocket, FromStr, FullOps, FusedIterator, Generator, Hash, Hasher, Index, IndexMut, InPlace, Int, Into, IntoCow, IntoInner, IntoIterator, IntoRawFd, IntoRawHandle, IntoRawSocket, IsMinusOne, IsZero, Iterator, JoinHandleExt, LargeInt, LowerExp, LowerHex, MetadataExt, Mul, MulAssign, Neg, Not, Octal, OpenOptionsExt, Ord, OsStrExt, OsStringExt, Packet, PartialEq, PartialOrd, Pattern, PermissionsExt, Place, Placer, Pointer, Product, Put, RangeArgument, RawFloat, Read, Rem, RemAssign, Seek, Shl, ShlAssign, Shr, ShrAssign, Sized, SliceConcatExt, SliceExt, SliceIndex, Stats, Step, StrExt, Sub, SubAssign, Sum, Sync, TDynBenchFn, Terminal, Termination, ToOwned, ToSocketAddrs, ToString, Try, TryFrom, TryInto, UnicodeStr, Unsize, UpperExp, UpperHex, WideInt, Write}
  , morekeywords=[2]{Send}  %
  , morekeywords=[3]{bool, char, f32, f64, i8, i16, i32, i64, isize, str, u8, u16, u32, u64, unit, usize, i128, u128}  %
  , morekeywords=[4]{Err, false, None, Ok, Some, true}  %
  , morekeywords=[3]{AccessError, Adddf3, AddI128, AddoI128, AddoU128, ADDRESS, ADDRESS64, addrinfo, ADDRINFOA, AddrParseError, Addsf3, AddU128, advice, aiocb, Alignment, AllocErr, AnonPipe, Answer, Arc, Args, ArgsInnerDebug, ArgsOs, Argument, Arguments, ArgumentV1, Ashldi3, Ashlti3, Ashrdi3, Ashrti3, AssertParamIsClone, AssertParamIsCopy, AssertParamIsEq, AssertUnwindSafe, AtomicBool, AtomicPtr, Attr, auxtype, auxv, BackPlace, BacktraceContext, Barrier, BarrierWaitResult, Bencher, BenchMode, BenchSamples, BinaryHeap, BinaryHeapPlace, blkcnt, blkcnt64, blksize, BOOL, boolean, BOOLEAN, BoolTrie, BorrowError, BorrowMutError, Bound, Box, bpf, BTreeMap, BTreeSet, Bucket, BucketState, Buf, BufReader, BufWriter, Builder, BuildHasherDefault, BY, BYTE, Bytes, CannotReallocInPlace, cc, Cell, Chain, CHAR, CharIndices, CharPredicateSearcher, Chars, CharSearcher, CharsError, CharSliceSearcher, CharTryFromError, Child, ChildPipes, ChildStderr, ChildStdin, ChildStdio, ChildStdout, Chunks, ChunksMut, ciovec, clock, clockid, Cloned, cmsgcred, cmsghdr, CodePoint, Color, ColorConfig, Command, CommandEnv, Component, Components, CONDITION, condvar, Condvar, CONSOLE, CONTEXT, Count, Cow, cpu, CRITICAL, CStr, CString, CStringArray, Cursor, Cycle, CycleIter, daddr, DebugList, DebugMap, DebugSet, DebugStruct, DebugTuple, Decimal, Decoded, DecodeUtf16, DecodeUtf16Error, DecodeUtf8, DefaultEnvKey, DefaultHasher, dev, device, Difference, Digit32, DIR, DirBuilder, dircookie, dirent, dirent64, DirEntry, Discriminant, DISPATCHER, Display, Divdf3, Divdi3, Divmoddi4, Divmodsi4, Divsf3, Divsi3, Divti3, dl, Dl, Dlmalloc, Dns, DnsAnswer, DnsQuery, dqblk, Drain, DrainFilter, Dtor, Duration, DwarfReader, DWORD, DWORDLONG, DynamicLibrary, Edge, EHAction, EHContext, Elf32, Elf64, Empty, EmptyBucket, EncodeUtf16, EncodeWide, Entry, EntryPlace, Enumerate, Env, epoll, errno, Error, ErrorKind, EscapeDebug, EscapeDefault, EscapeUnicode, event, Event, eventrwflags, eventtype, ExactChunks, ExactChunksMut, EXCEPTION, Excess, ExchangeHeapSingleton, exit, exitcode, ExitStatus, Failure, fd, fdflags, fdsflags, fdstat, ff, fflags, File, FILE, FileAttr, filedelta, FileDesc, FilePermissions, filesize, filestat, FILETIME, filetype, FileType, Filter, FilterMap, Fixdfdi, Fixdfsi, Fixdfti, Fixsfdi, Fixsfsi, Fixsfti, Fixunsdfdi, Fixunsdfsi, Fixunsdfti, Fixunssfdi, Fixunssfsi, Fixunssfti, Flag, FlatMap, Floatdidf, FLOATING, Floatsidf, Floatsisf, Floattidf, Floattisf, Floatundidf, Floatunsidf, Floatunsisf, Floatuntidf, Floatuntisf, flock, ForceResult, FormatSpec, Formatted, Formatter, Fp, FpCategory, fpos, fpos64, fpreg, fpregset, FPUControlWord, Frame, FromBytesWithNulError, FromUtf16Error, FromUtf8Error, FrontPlace, fsblkcnt, fsfilcnt, fsflags, fsid, fstore, fsword, FullBucket, FullBucketMut, FullDecoded, Fuse, GapThenFull, GeneratorState, gid, glob, glob64, GlobalDlmalloc, greg, group, GROUP, Guard, GUID, Handle, HANDLE, Handler, HashMap, HashSet, Heap, HINSTANCE, HMODULE, hostent, HRESULT, id, idtype, if, ifaddrs, IMAGEHLP, Immut, in, in6, Incoming, Infallible, Initializer, ino, ino64, inode, input, InsertResult, Inspect, Instant, int16, int32, int64, int8, integer, IntermediateBox, Internal, Intersection, intmax, IntoInnerError, IntoIter, IntoStringError, intptr, InvalidSequence, iovec, ip, IpAddr, ipc, Ipv4Addr, ipv6, Ipv6Addr, Ipv6MulticastScope, Iter, IterMut, itimerspec, itimerval, jail, JoinHandle, JoinPathsError, KDHELP64, kevent, kevent64, key, Key, Keys, KV, l4, LARGE, lastlog, launchpad, Layout, Lazy, lconv, Leaf, LeafOrInternal, Lines, LinesAny, LineWriter, linger, linkcount, LinkedList, load, locale, LocalKey, LocalKeyState, Location, lock, LockResult, loff, LONG, lookup, lookupflags, LookupHost, LPBOOL, LPBY, LPBYTE, LPCSTR, LPCVOID, LPCWSTR, LPDWORD, LPFILETIME, LPHANDLE, LPOVERLAPPED, LPPROCESS, LPPROGRESS, LPSECURITY, LPSTARTUPINFO, LPSTR, LPVOID, LPWCH, LPWIN32, LPWSADATA, LPWSAPROTOCOL, LPWSTR, Lshrdi3, Lshrti3, lwpid, M128A, mach, major, Map, mcontext, Metadata, Metric, MetricMap, mflags, minor, mmsghdr, Moddi3, mode, Modsi3, Modti3, MonitorMsg, MOUNT, mprot, mq, mqd, msflags, msghdr, msginfo, msglen, msgqnum, msqid, Muldf3, Mulodi4, Mulosi4, Muloti4, Mulsf3, Multi3, Mut, Mutex, MutexGuard, MyCollection, n16, NamePadding, NativeLibBoilerplate, nfds, nl, nlink, NodeRef, NoneError, NonNull, NonZero, nthreads, NulError, OccupiedEntry, off, off64, oflags, Once, OnceState, OpenOptions, Option, Options, OptRes, Ordering, OsStr, OsString, Output, OVERLAPPED, Owned, Packet, PanicInfo, Param, ParseBoolError, ParseCharError, ParseError, ParseFloatError, ParseIntError, ParseResult, Part, passwd, Path, PathBuf, PCONDITION, PCONSOLE, Peekable, PeekMut, Permissions, PhantomData, pid, Pipes, PlaceBack, PlaceFront, PLARGE, PoisonError, pollfd, PopResult, port, Position, Powidf2, Powisf2, Prefix, PrefixComponent, PrintFormat, proc, Process, PROCESS, processentry, protoent, PSRWLOCK, pthread, ptr, ptrdiff, PVECTORED, Queue, radvisory, RandomState, Range, RangeFrom, RangeFull, RangeInclusive, RangeMut, RangeTo, RangeToInclusive, RawBucket, RawFd, RawHandle, RawPthread, RawSocket, RawTable, RawVec, Rc, ReadDir, Receiver, recv, RecvError, RecvTimeoutError, ReentrantMutex, ReentrantMutexGuard, Ref, RefCell, RefMut, REPARSE, Repeat, Result, Rev, Reverse, riflags, rights, rlim, rlim64, rlimit, rlimit64, roflags, Root, RSplit, RSplitMut, RSplitN, RSplitNMut, RUNTIME, rusage, RwLock, RWLock, RwLockReadGuard, RwLockWriteGuard, sa, SafeHash, Scan, sched, scope, sdflags, SearchResult, SearchStep, SECURITY, SeekFrom, segment, Select, SelectionResult, sem, sembuf, send, Sender, SendError, servent, sf, Shared, shmatt, shmid, ShortReader, ShouldPanic, Shutdown, siflags, sigaction, SigAction, sigevent, sighandler, siginfo, Sign, signal, signalfd, SignalToken, sigset, sigval, Sink, SipHasher, SipHasher13, SipHasher24, size, SIZE, Skip, SkipWhile, Slice, SmallBoolTrie, sockaddr, SOCKADDR, sockcred, Socket, SOCKET, SocketAddr, SocketAddrV4, SocketAddrV6, socklen, speed, Splice, Split, SplitMut, SplitN, SplitNMut, SplitPaths, SplitWhitespace, spwd, SRWLOCK, ssize, stack, STACKFRAME64, StartResult, STARTUPINFO, stat, Stat, stat64, statfs, statfs64, StaticKey, statvfs, StatVfs, statvfs64, Stderr, StderrLock, StderrTerminal, Stdin, StdinLock, Stdio, StdioPipes, Stdout, StdoutLock, StdoutTerminal, StepBy, String, StripPrefixError, StrSearcher, subclockflags, Subdf3, SubI128, SuboI128, SuboU128, subrwflags, subscription, Subsf3, SubU128, Summary, suseconds, SYMBOL, SYMBOLIC, SymmetricDifference, SyncSender, sysinfo, System, SystemTime, SystemTimeError, Take, TakeWhile, tcb, tcflag, TcpListener, TcpStream, TempDir, TermInfo, TerminfoTerminal, termios, termios2, TestDesc, TestDescAndFn, TestEvent, TestFn, TestName, TestOpts, TestResult, Thread, threadattr, threadentry, ThreadId, tid, time, time64, timespec, TimeSpec, timestamp, timeval, timeval32, timezone, tm, tms, ToLowercase, ToUppercase, TraitObject, TryFromIntError, TryFromSliceError, TryIter, TryLockError, TryLockResult, TryRecvError, TrySendError, TypeId, U64x2, ucontext, ucred, Udivdi3, Udivmoddi4, Udivmodsi4, Udivmodti4, Udivsi3, Udivti3, UdpSocket, uid, UINT, uint16, uint32, uint64, uint8, uintmax, uintptr, ulflags, ULONG, ULONGLONG, Umoddi3, Umodsi3, Umodti3, UnicodeVersion, Union, Unique, UnixDatagram, UnixListener, UnixStream, Unpacked, UnsafeCell, UNWIND, UpgradeResult, useconds, user, userdata, USHORT, Utf16Encoder, Utf8Error, Utf8Lossy, Utf8LossyChunk, Utf8LossyChunksIter, utimbuf, utmp, utmpx, utsname, uuid, VacantEntry, Values, ValuesMut, VarError, Variables, Vars, VarsOs, Vec, VecDeque, vm, Void, WaitTimeoutResult, WaitToken, wchar, WCHAR, Weak, whence, WIN32, WinConsole, Windows, WindowsEnvKey, winsize, WORD, Wrapping, wrlen, WSADATA, WSAPROTOCOL, WSAPROTOCOLCHAIN, Wtf8, Wtf8Buf, Wtf8CodePoints, xsw, xucred, Zip, zx}
  , morekeywords=[5]{assert!, assert_eq!, assert_ne!, cfg!, column!, compile_error!, concat!, concat_idents!, debug_assert!, debug_assert_eq!, debug_assert_ne!, env!, eprint!, eprintln!, file!, format!, format_args!, include!, include_bytes!, include_str!, line!, module_path!, option_env!, panic!, print!, println!, select!, stringify!, thread_local!, try!, unimplemented!, unreachable!, vec!, write!, writeln!}  %
}

\lstdefinestyle{coloredRust}
  {
    language=Rust
    , style = common
    , keywordstyle=\bfseries%
    , keywordstyle=[2]\color[rgb]{0.75, 0, 0}%
    , keywordstyle=[3]\color[rgb]{0, 0.5, 0}%
    , keywordstyle=[4]\color[rgb]{0, 0.5, 0}%
    , keywordstyle=[5]\color[rgb]{0, 0, 0.75}%
    , morecomment=[l]{\//},
  }

\lstnewenvironment{coderust}
  {\lstset{style=coloredRust}}
  {}

\lstnewenvironment{coderustfloat}
  {\lstset{style=coloredRust, belowskip=-0.5\baselineskip,float=ht}}
  {}

\lstdefinelanguage{Vrs}{%
  sensitive%
, morecomment=[l]{//}%
, morecomment=[s]{/*}{*/}%
, moredelim=[s][{\itshape\color[rgb]{0,0,0.75}}]{\#[}{]}%
, morestring=[b]{"}%
, alsodigit={}%
, alsoother={}%
, alsoletter={!}%
, morekeywords={break, continue, else, for, if, in, not, loop, match, return, while}  %
, morekeywords={as, const, let, move, mut, ref, static}  %
, morekeywords={unit, segment, staticmap, mem, reg, mmio, fn, requires, ensures, forall, abstract, synth, exists, enum, dom}  %
, morekeywords={Register, Memory, MMIO, CPURegister, StateDef, InterfaceDef}  %
, morekeywords=[2]{Layout, ReadActions, WriteActions, maps}
, morekeywords=[3]{inaddr, outaddr, flags,size, int, nat, addr, vaddr, paddr, bool, mode}  %
, morekeywords=[4]{true, false}  %
, morekeywords=[4]{state, interface, mapdef,inbitwidth, outbitwidth, State}  %
}

\lstdefinestyle{coloredvrs}
{
  language=Vrs
  , style = common
  , keywordstyle=\bfseries%
  , keywordstyle=[2]\color[rgb]{0.75, 0, 0}%
  , keywordstyle=[3]\color[rgb]{0, 0.5, 0}%
  , keywordstyle=[4]\color[rgb]{0, 0.5, 0}%
  , keywordstyle=[5]\color[rgb]{0, 0, 0.75}%
  , morecomment=[l]{\//}
  , morecomment=[l]{\#}
}

\lstnewenvironment{codevrs}
  {\lstset{style=coloredvrs}}
  {}

\newcommand{\inlinecodevrs}[1]{\lstinline[style=coloredVrs, basicstyle=\lstinlinebasicstyle, columns=fullflexible]{#1}}

\lstnewenvironment{codevrsfloat}[1]
  {\lstset{style=coloredvrs,float=t, belowskip=-0.5\baselineskip, #1}}
  {}

\lstnewenvironment{codevrsfloat*}[1]
  {\lstset{style=coloredvrs,float=*, belowskip=-0.5\baselineskip, #1}}
  {}

\lstdefinelanguage{z3}{
  sensitive=true,
  alsoletter={\-, \.},
  comment=[l]{;},
  keywords=[1]{
  apply, assert, assert-soft, check-sat, check-sat-using, compute-interpolant,
  declare-const, declare-datatypes, declare-fun, declare-map, declare-rel,
  declare-sort, declare-tactic, define-sort, display, echo, eval, exit,
  fixedpoint-pop, fixedpoint-push, get-assertions, get-assignment, get-info, get-
  interpolant, get-model, get-option, get-proof, get-unsat-core, get-user-tactics,
  get-value, help, help-tactic, labels, maximize, minimize, pop, push, query,
  reset, rule, set-info, set-logic, set-option, simplify, forall, exists, let,
  define-fun
  },
  morekeywords=[2]{
  check-sat-using, declare-var, declare-rel, rule, query, set-predicate-
  representation, maximize, minimize, assert-soft, assert-weighted, compute-
  interpolant
  },
  morekeywords=[3]{and, bvuge, bvule, bvand, bvlshr, =>, =, !},
  morekeywords=[4]{BitVec, VAddr_t, Bool, Num_t, PAddr_t, Model_t, Size_t, Flags_t},
}

\lstdefinestyle{coloredz3}
{
  language=z3
  , style = common
  , keywordstyle=\bfseries%
  , keywordstyle=[2]\color[rgb]{0.75, 0, 0}%
  , keywordstyle=[3]\color[rgb]{0.75, 0, 0}%
  , keywordstyle=[4]\color[rgb]{0, 0.75, 0}%
  }

\lstnewenvironment{codez3}
  {\lstset{style=coloredz3}}
  {}

\lstnewenvironment{codez3float}[1]
  {\lstset{style=coloredz3,float=ht, belowskip=-0.5\baselineskip, #1}}
  {}

\ifisdraft

\newcommand{\nbc}[3]{
    {{\colorbox{#3}{\bfseries\sffamily\scriptsize\textcolor{white}{#1}}}}
    {\textcolor{#3}{\sf\small$\blacktriangleright~$\textit{#2}$~\blacktriangleleft$}}}

\else
\newcommand{\nbc}[3]{}
\fi

\newcommand{\system}{\emph{Velosiraptor}\xspace}

\newcommand{\osmap}{\inlinecodevrs{addmap}\xspace}
\newcommand{\osprot}{\inlinecodevrs{setperms}\xspace}
\newcommand{\osunmap}{\inlinecodevrs{delmap}\xspace}
\newcommand{\osfuns}{\osmap, \osprot, and \osunmap\xspace}

\newcommand{\hwmap}{map\xspace}
\newcommand{\hwmapfn}{\inlinecodevrs{map()}\xspace}
\newcommand{\hwmapping}{mapping\xspace}
\newcommand{\Hwmapping}{Mapping\xspace}

\newcommand{\hwuint}{mapping unit\xspace}
\newcommand{\hwunit}{mapping unit\xspace}
\newcommand{\HwUnit}{Mapping Unit\xspace}

\copyrightyear{2025}
\acmYear{2025}
\setcopyright{cc}
\setcctype{by}
\acmConference[ASPLOS '25]{Proceedings of the 30th ACM International Conference on Architectural Support for Programming Languages and Operating Systems, Volume 2}{March 30-April 3, 2025}{Rotterdam, Netherlands}
\acmBooktitle{Proceedings of the 30th ACM International Conference on Architectural Support for Programming Languages and Operating Systems, Volume 2 (ASPLOS '25), March 30-April 3, 2025, Rotterdam, Netherlands}
\acmDOI{10.1145/3676641.3711998}
\acmISBN{979-8-4007-1079-7/25/03}


\begin{document}

\title{\system: Code Synthesis for Memory Translation}
\acmSubmissionID{182}

\author{Reto Achermann}
\orcid{0000-0003-3263-7236}
\affiliation{%
\institution{University of British Columbia}
\city{Vancouver, BC}
\country{Canada}
}

\definecolor{greencolor}{rgb}{0.11,0.63,0.14}

\author{Em Chu}
\orcid{0009-0000-6673-4945}
\affiliation{%
\institution{JuliaHub}
\city{Boston, MA}
\country{USA}
}
\authornote{Work done while at The University of British Columbia.}

\definecolor{purplecolor}{rgb}{0.56,0.10,0.63}

\author{Ryan Mehri}
\orcid{0009-0007-0559-4820}
\affiliation{%
\institution{Replit}
\city{San Francisco, CA}
\country{USA}
}
\authornotemark[1]

\definecolor{purplecolor}{rgb}{0.56,0.10,0.63}

\author{Ilias Karimalis}
\orcid{0009-0004-6594-0359}
\affiliation{%
\institution{University of British Columbia}
\city{Vancouver, BC}
\country{Canada}
}

\definecolor{purplecolor}{rgb}{0.56,0.10,0.63}

\author{Margo Seltzer}
\orcid{0000-0002-2165-4658}
\affiliation{%
\institution{University of British Columbia}
\city{Vancouver, BC}
\country{Canada}
}

\definecolor{bluecolor}{rgb}{0.11,0.32,1.00}

\renewcommand{\shortauthors}{Reto Achermann, Em Chu, Ryan Mehri, Ilias Karimalis, and Margo
Seltzer} 

\begin{abstract}

Security is among the top concerns of operating system (OS) developers.
A secure runtime environment relies on the OS to correctly configure the memory hardware on which it
runs.
This is mission-critical as it provides essential security-relevant features and abstractions that
ensure the integrity and isolation of untrusted applications running alongside each other.
Configuring a platform's memory hardware is not a one-off effort as designers constantly
develop new mechanisms for translation and protection with different features and means of
configuration.
Adapting the OS code to the new hardware is not only a manual, repetitive and time-consuming task,
it may also introduce subtle, but security-critical bugs that break security and isolation
guarantees.

We present \system,
a system that automatically generates correct, low-level OS code that programs
the memory hardware of a machine.
\system leverages software synthesis techniques and exploits the domain specificity of the problem
to make the synthesis process efficient.
With \system, developers write only a high-level description of the memory hardware's mapping
behavior and OS environment.

The \system toolchain transforms this specification into a verified implementation that can be
linked directly with the rest of the operating system.
Incorporating the OS environment into this process allows porting an OS to new hardware platforms
without worrying about writing code to configure the memory hardware.
We can also use the same specification to generate hardware components.
This enables research in new translation mechanisms, freeing up OS developers from
manually writing OS code.
 \end{abstract}

\maketitle %

\section{Introduction}
\label{sec:introduction}

\remove{use the $remove$  command to indicate the text that can be removed}

\remove{System software is responsible for providing a secure runtime environment for applications.
Enforcing isolation between applications, e.g., processes, containers or virtual machines, is,
therefore, mission-critical for operating systems and hypervisors.
Thus, the security of the runtime environment is one of the primary concerns of OS
developers~\cite{Mahfud:2023:SLR}.
}

Today's computer systems contain many memory protection units (MPUs), memory management units
(MMUs), and other memory hardware or features that mediate access to memory resources.
A single memory access may traverse multiple hardware units until it reaches its destination.
Generally, those hardware components provide \emph{\hwmapping} functionality that can be split up
into \emph{protection} and \emph{translation}~\cite{Alam:2017:DVMT}.

Protection determines if a specific memory access is permitted.
This involves checking whether the input address is valid and the access mode or source of the
memory access is permissible given the protection attributes (e.g., write, privilege level, or a
specific device).
Translation is the process of transforming the address of a memory access.
\Hwmapping is a partial function from an input address to an output address, i.e., the \hwmapping
attempt can fail due to an invalid address or insufficient permissions.

The OS must program \emph{all} memory hardware units of the machine correctly, including the
processor's MMUs, the System MMU, and other memory translation and protection units isolating
individual hardware components to ensure the security of applications~\cite{Dorr:2021:MCA,
Fiedler:2023:SOS, Fiedler:2023:PHDF}.
This means that OS developers must have precise knowledge of all the memory hardware and all its
subtle differences in their features and configuration options.
For example, the OS must set the right permission flags or memory attributes in the page table
entry.
Therefore, the developer has to \emph{manually} write the low-level OS code that programs the \hwmapping
hardware.

Unfortunately, writing code to configure memory hardware is not a one-time effort and can take
weeks~\cite{Rashid:1987:MIVA}.
Modern platforms have many different memory hardware units mediating access to and from various
chip subsystems.
Those units have a diverse set of features and means of configuration that the developer needs to
understand.
Thus, the developer has to spend a significant amount of time repeating the same task: reading the
hardware manual and then manually writing and debugging configuration code -- a process that
easily introduces subtle bugs \cite{CVE-2014-9888,CVE-2017-16994, Morgan:2016:BIP,
Morgan:2018:IPIO,Markuze:2016:TIP,Markettos:2019:TEV, Huang:2016:ESL} amplified by the repetition.
Any error in the configuration sequence can break system security, allowing applications to
inadvertently corrupt data or obtain unauthorized access to resources of other applications.

We build upon previous work~\cite{hotos} and present \system\footnote{\emph{\underline{V}e\underline{l}o\underline{si}raptor} is a wordplay on VLSI -- synthesis for hardware logic.}, a system that automatically generates
\emph{correct} memory configuration code.
In \system, developers no longer need to manually and tediously write this code. Instead, they
obtain a behavioral specification of the memory hardware as part of the technical hardware manual,
or they write a specification that closely matches the description in the manual.
This behavioral specification defines \emph{how} the memory hardware performs address \hwmapping.
The \system toolchain then constructs a model of the memory hardware and uses software synthesis
to automatically generate the OS code that configures the memory hardware.
The behavioral specification needs to be written once per \hwunit.
\system then tailors the generated code to a target environment by using an OS specification that
describes, for instance, the memory allocation functions or capability invocations.

\system leverages domain-specific knowledge to make software synthesis efficient.
\system uses divide-and-conquer to reduce the search space of candidate programs by orders of
magnitude, resulting in a negligible kernel compilation time overhead.
This makes \system a viable approach to automatically generating correct, low-level, memory
hardware configuration code.

\system also enables research into new memory translation and protection mechanisms by generating a
simulated hardware component from the same behavioral specification.
This lets developers evaluate the utility of proposed, but non-existent memory hardware
~\cite{Basu:2023:EVM, Gosakan:2023:MPBT, Landgraf:2024:RVM}.
Researchers no longer have to build a hardware implementation to evaluate the efficacy of a newly
proposed translation and/or protection mechanism.
Currently, \system supports the Arm Fast Models~\cite{Arm:2024:FastModels}, a functionally accurate
platform simulator.

We make the following contributions:
\begin{itemize}
    \item A methodology to efficiently synthesize code interfacing with memory hardware (\autoref{sec:synthesis} and \autoref{sec:eval:scaling}),
    \item a domain-specific language to concisely  express the behavior of memory hardware (\autoref{sec:language}),
    \item the implementation of \system in Rust that uses Z3 for program verification (\autoref{sec:implementation}),
    \item a backend for \system that generates memory hardware modules for the Arm
    Fast Models (\autoref{sec:impl:hwgen}),
    \item a quantitative evaluation showing the effectiveness of the search space reduction and
     the efficiency of the synthesis process (\autoref{sec:eval:optimizations}), and
    \item a qualitative demonstration of the functionality of the generated memory hardware
      module and the adaption of the generated code to the OS environment (\autoref{sec:eval:os}).
\end{itemize}

\section{Modeling Memory Hardware Behavior}
\label{sec:methodology}

Modern memory hardware is diverse in both the features it provides (e.g., means of translation and
protection) and how the hardware is programmed (e.g., registers and translation descriptors).
For example, both Arm and x86 use an in-memory page table to define virtual-to-physical address
translation.
They offer similar functionality but have different layouts.
The Xeon Phi co-processor~\cite{intel:xeonphi} uses an array of registers to control memory accesses
to the host memory.
Despite their heterogeneity, they share many common characteristics that allow for a common,
abstract model that expresses their \hwmapping behavior.
For example, determining if the hardware \hwmap{}s a given input depends on state that is stored
either in registers or an in-memory data structure.
We use the term \emph{\hwmap} to encompass both translation and access control.
From the behavioral model, we define an operational model that captures the concrete state and
software visible interface.
Finally, we define the basic building block and composition rules with well-defined semantics.

\subsection{Behavioral Model}
\label{sec:model:behavioral}

Abstractly, all memory hardware units either successfully \hwmap an address or raise an exception.
Thus, we can model \hwmap behavior as a partial function from an input address, an access mode
(e.g., read, write, execute, \ldots), and the memory hardware state (e.g., register contents)
to an output address.
\[
  \texttt{\hwmap} :: inaddr \rightarrow mode \rightarrow state \rightharpoonup outaddr
\]
Input and output addresses may be ordinary memory addresses or they may also contain source
information, e.g., which core or device initiated the memory access.
The set of possible access modes can be defined per \hwunit, e.g., to include processor modes such
as user and supervisor mode.
The state includes everything that can influence translation (e.g., registers and in-memory data
structures such as page tables).
The \hwmapfn function produces an output address if and only if the state permits access to the
input address under the current access mode.
Mathematically, the \hwmapfn function produces an output address if and only if the input values are
part of the function's domain: $(inaddr, mode, state) \in dom(\texttt{\hwmap})$.
For example, the corresponding entry in the page table must be valid and have the right permission
bits set, otherwise, an exception is raised.

We currently focus on the \hwmapping behavior and leave triggering a specific exception class (e.g.,
the distinction between a translation fault and a write fault) for future work.

\begin{figure}[ht!]
  \includegraphics{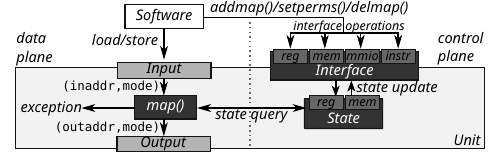}
  \caption{The behavioral model of a \hwmapping unit.}
  \label{fig:model}
\end{figure}

\subsection{Operational Model}

\autoref{fig:model} depicts the operational model of a \emph{\hwuint} (we use this term to refer to
any memory hardware component that mediates memory accesses through address translation or access
control).
The operational model of a \hwuint consists of three main components: \hwmapfn, state, and interface
(shown by the dark boxes in \autoref{fig:model}).
It resembles a state machine where each state has a well-defined \hwmapping behavior.
The state consists of registers and memory locations, and the OS triggers state transitions by
reading or writing special or MMIO registers or memory locations or executing specific instructions.
Similar to a network switch, the operational model consists of a data plane and a control plane, but
instead of processing network packets, it processes memory accesses, i.e., load or
store instructions or DMA transfers.

\subsubsection{Synthesis Target Specifications}

The memory management subsystem must provide three functions: \osmap adds a new mapping from an
input address range to an output address range with specified permissions, \osprot changes the
access permissions of an address range, and \osunmap removes a mapping.
\autoref{code:hlfn} shows the corresponding function signatures, which serve as the specifications
for our synthesis targets.
The predicate \inlinecodevrs{allows(f, m)} defines whether the access mode is permitted given the
supplied permission flags.
The synthesis task is then to find an implementation by finding a sequence of state machine
transitions (i.e., interface interactions) that produces the goal state defined as post-conditions.
Note, that there is an implicit post-condition ensuring that the behavior outside of the specified
address range is not changed.
\begin{codevrsfloat}{label=code:hlfn, caption={Synthesis target function specifications where \hwmapfn refers to the \hwmap-function defined in \autoref{sec:model:behavioral}}}
fn addmap(va:vaddr, sz:size, f:flags, pa:paddr, (*@ \label{line:hlfn:map} @*)
          s:State) -> (s':State)
  ensures forall v:vaddr, m:mode ::
    inrange(va,sz,v) ==> (translate(s',v) == pa+(v-va)
      && (allows(f, m) <==> (v,m,s') in dom(map)))

fn setperms(va:vaddr, sz:size, f:flags, pa:paddr (*@ \label{line:hlfn:protect} @*)
            s:State) -> (s':State)
  ensures forall v:vaddr, m:mode ::
    inrange(va,sz,v) ==> (translate(s',v)==translate(s,v)
      && (allows(f, m) <==> (v,m,s') in dom(map)))

fn delmap(va:vaddr, sz:size, f:flags, pa:paddr (*@ \label{line:hlfn:unmap} @*)
          s:State) -> (s':State)
  ensures forall v:vaddr, m:mode ::
    inrange(va,sz,v) ==> (v,m,s') not in dom(map)
\end{codevrsfloat}

\subsubsection{Control Plane}

We model the control interface as a state machine.
The state of a \hwuint is a sequence of bits, stored in registers or memory locations and
includes all relevant parts that define the \hwmapping behavior.
For example, on x86, the page table defines the virtual-to-physical address translation, whereas the
CR4 (control register 4) defines whether the \hwmapping is enabled~\cite{intel:sdm}.
The control interface of the \hwuint defines the transitions of the state machine.
Each transition is atomic, takes a sequence of arguments and an input state, and produces a
well-defined, new state.
Examples include updating a page table entry, writing the translation base register, or executing
the \texttt{tlbwr} instruction on the MIPS R4700~\cite{mips}.
Formally, this is:
\[
  \texttt{transition} :: Seq \text{<} arg \text{>} \rightarrow state \rightarrow state
\]
We expect that most transitions take zero or one machine-word-sized argument, representing the
payload of load, store or other instructions.
A special \texttt{init}-transition defines the bit pattern of the initial state.
The control interface then consists of a set of transitions and the \texttt{init} transition.

In the model, the OS cannot access the state of the \hwunit directly.
Instead, the OS interacts with the control interface -- the \hwunit's public API.
To change a unit's \hwmapping behavior, the OS puts the state machine into the desired state by
applying the right sequence of atomic transitions.
This notion becomes important in \autoref{sec:synth:programs} as the available state
transitions correspond to the \emph{grammar} of the programs that are needed to configure the
\hwmapping behavior.

\subsubsection{Data Plane}

Recall, that the \hwmapping has two parts: translation and access control.
A boolean expression distinguishes whether the \hwmapfn function produces an error or an exception
(we do not model different exception types).

We express access control checks as boolean predicates in conjunctive normal form.
The conjuncts are boolean functions defined over the input address, access mode, and state.
Output address computation occurs if, and only if, all the conjuncts evaluate to true; otherwise the
unit raises an exception.
Formally, we express this as:
\[
  (addr, mode, state) \in dom(\texttt{\hwmap} ) \Leftrightarrow  \bigwedge_{p\in P} p(addr, mode, state)
\]
If the input address is outside of the valid range or the access mode is not permitted, at
least one of the conjuncts evaluates to false, rendering the \hwmapfn function undefined for the
input arguments.
The formalization using the conjuncts helps us state the problem of synthesizing the correct
configuration program for the \hwuint (\autoref{sec:synth:reduction}).

\subsection{Composition of Building Blocks}
\label{sec:methodology:buildingblocks}

Given the diversity in memory hardware, we designed a methodology of basic building blocks and
composition rules from which we compose specifications hierarchically.
This modular approach makes synthesis easier and more amenable to optimization~\cite{Hu:2023:TPOS}.
After analyzing more than 12 different memory hardware units, we were pleasantly surprised that we
\emph{identified only one basic building block and two composition rules from which we could specify
all of them.}

\subsubsection{The Basic Building Block}

The basic building block is a \emph{segment} that \hwmap{}s a contiguous input address range to a
contiguous output address range.
The output address range might be the ultimate result of a map or it might be an intermediate result,
providing the address of the next unit in a structure such as the x86 tree-based page table.
The segment has both a state and a control interface.
The input or output address range, i.e., the base address and size, may be either fixed (i.e.,
hardwired to a certain value) or configurable (e.g., computed from the state).
For example, the Xeon Phi entry system memory page table~\cite{intel:xeonphi} maps a fixed input
address range of 16 GB.
In contrast, the x86 segment descriptor~\cite{intel:sdm} allows mappings of different sizes, and
x86 page directory entries may map to the next-level page table encoding the tree structure.

\subsubsection{Combining Basic Building Blocks}

Developers combine segments using the composition rules.
Each rule has well-defined semantics and dictates constraints on the composite state arising from
the combined states of the basic building blocks.
Rules do not have state or control interfaces.

\myparagraph{Enumeration}
Enumerations provide a way to create a structure akin to a C union from two or more segments.
Each variant of the enumeration is a segment that has the same state bits, but it may interpret
those state bits differently.
For example, a page directory entry can map a large frame or a page table, depending on whether the
page-size bit is set.
Variants must be non-conflicting, i.e., the state must unambiguously identify which segment
description provides the right interpretation of the state.

\myparagraph{StaticMap}
A static map defines a table-like construct composed of a collection of entries, each of which
covers an independent, non-overlapping, and contiguous address range.
Each entry in a static map may be either a segment or an enumeration, but not another static map.
In principle, \system supports entries of different types, but we have not encountered this in the
\hwunit{}s we analyzed.
Some part of the input address must specify which table entry contains the mapping for the input address.
Like the enumeration, the map links the state of the referenced segments together, adding
constraints on the combined state.

\subsubsection{Composition Example}

\begin{figure}
  \begin{center}
    \begin{scriptsize}
  \includegraphics{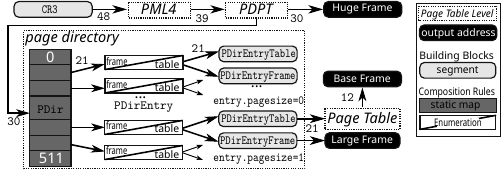}

\begin{verbatim}
  PDirEntryTable -> PageTable;        PDirEntryFrame -> LargeFrame;
  PDirEntry      = [ PDirEntryTable | PDirEntryFrame ];
  PDir           = [ PDirEntry; 512 ];
\end{verbatim}%
\end{scriptsize}
\end{center}
  \caption{Decomposition of the four-level x86\_64 page table into basic building blocks and composition rules.}
  \label{fig:composition}
\end{figure}

We use a simplified representation of the four-level x86\_64 virtual address translation as an
illustrative example (we exclude segmentation and the TLB for simplicity).
From the manual~\cite{intel:sdm}, we learn that address \hwmapping accesses a four-level radix tree of
in-memory descriptor tables, a translation base register (CR3), and registers that control translation
features such as physical address extension and whether the MMU is enabled (omitted in
\autoref{fig:composition}).
The OS configures the \hwmapping behavior by writing specific bit patterns into the registers or
page table entries.

\autoref{fig:composition} shows the decomposition of the x86\_64 address translation into \system
building blocks with the page directory shown in detail.
We describe the behavior of these components by following a page table walk.
A walk begins with access to the base register (CR3) and then traverses through a four-level radix
tree (PML4, PDPT, PDIR, PTable) as indicated by the dotted boxes of \autoref{fig:composition}.
Having identified the structures we need, we next recursively decompose the components into building
blocks: CR3 cannot be split up further, so we model it as a segment unit.
Each level in the tree is an array of entries each mapping a fixed, non-overlapping part of the
input address range.
The page directory allows the mapping of either a page table or a large page (differentiated by the
page size bit).
Therefore, we use the enumeration rule to combine two segments -- one mapping a frame
(\inlinecodevrs{PDirEntryFrame}) and the other mapping the page table
(\inlinecodevrs{PDirEntryTable}) -- to form a page directory entry (\inlinecodevrs{PDirEntry}).
Finally, the enumerations are then combined using the static map composition rule to form the table.

\subsection{\HwUnit Type Hierarchy}

From the composition example above, we see that expressing the \hwuint as a combination of segment
units and composition rules is similar to a type or class hierarchy.
Indeed, each of the defined segment units and rule applications defines a type.
The specification defines how the types are instantiated and how they are referenced by the units.
Thus, from the specification of the \hwuint, we can obtain the type hierarchy.
Further, by knowing the type hierarchy, we can extract the structure of in-memory data structures
such as the two descriptor tables used in the example.
Finally, using the types we can reuse the basic building blocks in other \hwuint descriptions.

\section{Specification Language}
\label{sec:language}

Developers write two specifications in the \system Specification Language:
The first specification describes the \hwmapping behavior of the hardware using intuitive language
constructs for the concepts outlined in the previous section.
The second specification describes the runtime environment provided by the OS, e.g., memory
allocation functions.

We explain the language constructs and how they relate to the behavioral model using the X86\_64
page directory from \autoref{fig:composition} and show the slightly simplified specification in
\autoref{code:language}.
We omit parts of the state and interface description and \osprot and \osunmap for readability.

\begin{codevrsfloat}{label=code:language, caption={Simplified x86\_64 Page Directory Specification}, float}
// PDir table as static map
staticmap PDir(base: paddr) (*@ \label{line:unit:pdir} @*)
  {  maps [ PDirEntry(base + i*PTABLE_ENTRY_SIZE)
            for i in 0..PTABLE_NUM_ENTRIES ];  } (*@ \label{line:unit:mapdef} @*)
// PDirEntry with two variants
enum PDirEntry(base: paddr) (*@ \label{line:unit:pdire} @*)
  { PDirEntryTable(base), PDirEntryFrame(base) }
// PDir entry mapping a page table (derived)
segment PDirEntryTable(base: paddr) : PTableDesc {(*@ \label{line:unit:pdire:table} @*)
  synth fn addmap(va:vaddr,sz:size,flgs:flags,pa:PTable)(*@ \label{line:unit:pdire:map} @*)
    requires  pa & (BASE_PAGE_SIZE - 1) // ...
}
// PDir entry mapping a large frame
segment PDirEntryFrame(base: paddr) {(*@ \label{line:unit:pdire:page} @*)
  state(base: paddr) { (*@ \label{line:unit:state} @*)
    mem entry [ base, 0, 8] {  (*@ \label{line:unit:state:field} @*)
       0 ..  1 present, (*@ \label{line:unit:state:bitslice} @*)
       1 ..  2 writable, // bits 2..7 removed for space
       7 ..  8 pagesize, // bits 8..21 removed for space
      21 .. 48 address,
  } }
  interface(base: paddr) { (*@ \label{line:unit:interface} @*)
    mem entry [ base, 0, 8 ] {
      Layout { /* same as state */ }
      WriteActions { interface.entry -> state.entry; } (*@ \label{line:unit:writeaction} @*)
      ReadActions  { state.entry -> interface.entry; } (*@ \label{line:unit:readaction} @*)
  } }
  #[pred]
  fn valid() -> bool (*@ \label{line:unit:fn:valid} @*)
  { state.entry.present == 1 }
  #[pred]
  fn is_writable(flgs:flags) -> bool (*@ \label{line:unit:fn:is_writable} @*)
  { flgs.writable ==> state.entry.writable }

  fn translate(va:vaddr) -> paddr(*@ \label{line:unit:fn:translate} @*)
    requires state.entry.pagesize == 1  (*@ \label{line:unit:pagesize} @*)
  { va + (state.entry.address << LARGE_PAGE_BITS) }

  synth fn addmap(va:vaddr,sz:size,flgs:flags,pa:paddr) (*@ \label{line:unit:fn:map} @*)
    requires va == 0 && sz == LARGE_PAGE_SIZE
    requires pa & (LARGE_PAGE_SIZE - 1) == 0
} // end of PDirEntryFrame
\end{codevrsfloat}

\subsection{Language Overview}

\autoref{code:language} shows the definition of two basic building blocks and two composition rules
that correspond to the x86\_64 page directory shown in \autoref{fig:composition}.
Line~\ref{line:unit:pdir} defines the page directory (\inlinecodevrs{PDir}) as a static map.
The \inlinecodevrs{base} parameter is the address where the \inlinecodevrs{PDir} is allocated in memory.
We use list-comprehension notation to express that the \inlinecodevrs{PDir} is a table of
\inlinecodevrs{PTABLE_NUM_ENTRIES=512} entries.
We instantiate each entry at an address relative to the \inlinecodevrs{base} parameter to associate
individual entries with the full table.

An entry in the \inlinecodevrs{PDir} has two variants, so we use the \inlinecodevrs{enum} keyword
to define an enumeration (\inlinecodevrs{PDirEntry}) on line \ref{line:unit:pdire}.
The enumeration lists the variants and constrains the variant's parameters by binding them to the
corresponding parameter of the enumeration (\inlinecodevrs{base} parameter).
The first variant maps a page table that we express as a configurable \inlinecodevrs{segment}.
In x86, table entries mapping other tables have similar representations.
To avoid duplication, \system provides a notion of inheritance.
On line~\ref{line:unit:pdire:table}, we specify the \inlinecodevrs{PDirEntryTable} by \emph{deriving}
it from the generic page table descriptor \inlinecodevrs{PTableDesc} (omitted) and specialize the \osmap
function (Line~\ref{line:unit:pdire:map}).
Note the type of the \inlinecodevrs{pa} parameter here defines the type of the next unit, i.e.,
\inlinecodevrs{PTable}, effectively encoding the page table tree structure.
The second variant maps a large frame.
Thus, on line \ref{line:unit:pdire:page} we declare a segment unit (\inlinecodevrs{PDirEntryTable}).

\subsection{Specification Details}

The \system language has specific constructs to express the concepts of the operational model
(\autoref{sec:methodology}) and allows developers to intuitively specify the \hwmapping behavior.

\subsubsection{State Definitions}

Recall, that the state is just a sequence of bits in the operational model.
The state definition (line~\ref{line:unit:state}) assigns names to specific groups of bits of the
state.
The state definition can be interpreted as declaring private data members of a class.
Logically, the state is divided into a set of \emph{fields} that correspond to individual registers
or words in memory.
Register fields (declared with the \inlinecodevrs{reg} keyword) have just a size, while in-memory
fields (\inlinecodevrs{mem} keyword) also have an address parameter and an offset.
For example, line~\ref{line:unit:state:field} defines an 8-byte in-memory location, named
\inlinecodevrs{entry}, at offset \inlinecodevrs{0} from \inlinecodevrs{base}.
Registers or memory words may further contain groups of bits with specific semantic meanings.
Developers can define \emph{bit slices} \inlinecodevrs{a..b}, which assign names to consecutive ranges
of bits from  \inlinecodevrs{a} to  \inlinecodevrs{b} (exclusive).
Line \autoref{line:unit:state:bitslice} assigns bit slice \inlinecodevrs{0..1} (i.e., bit 0) the
name \inlinecodevrs{present}, so it can be referenced by a semantically meaningful name.

\subsubsection{Interface Definitions}

The control interface definition (line~\ref{line:unit:interface}) specifies the software-visible API
and follows a similar pattern to the state definition by assigning names to parts of the interface.
Recall, that the operational model does not permit directly accessing the state of the unit.

The interface is divided into a set of fields each of which may have a layout (i.e., bit slices) and
a set of actions (i.e., \inlinecodevrs{ReadActions} and \inlinecodevrs{WriteActions}) that specify
how the state is updated.
Each field is either: 1) an in-memory location (\inlinecodevrs{mem}) that is an overlay of an
in-memory state field, 2) a memory-mapped register (\inlinecodevrs{mmio}), 3) a non-memory-mapped
register (\inlinecodevrs{reg}), or 4) an instruction (\inlinecodevrs{instr}).

Action blocks contain a list of data transfer actions that atomically execute whenever the interface
field is accessed.
Lines \ref{line:unit:writeaction} and \ref{line:unit:readaction} show the data transfer between the
interface and the state where the \inlinecodevrs{state} and \inlinecodevrs{interface} variables
refer to the current unit instance.
Note that for fields defined with \inlinecodevrs{mem}, the actions are restricted to the corresponding
\inlinecodevrs{mem} state fields.
Register-backed interface fields may trigger multiple non-conflicting updates to the entire state

As an optimization, developers may omit the layout and action blocks if there is a state field with
the same name and kind.
In that case, the interface field acts as an overlay and inherits the same layout as the state field
and the default actions will simply pass through all reads and writes.

\subsubsection{Methods}

Methods, defined with the keyword \inlinecodevrs{fn}, contain expressions to compute a value and
specify the \hwmapping behavior.
They can access the state directly through the \inlinecodevrs{state} variable referring to
the state of the current unit instance (such as the \texttt{this} pointer).
Methods can further specify \inlinecodevrs{requires} clauses that express pre-conditions of the
method, which serve as constraints in the synthesis process.
Methods effectively create named expressions that can call each other (we disallow loops and
recursion).
There are three special method classes.

\myparagraph{\hwmapping Predicates}
As outlined in the behavioral model, the \hwmapping function is defined only if all conjuncts are
true.
Developers define such conjuncts by tagging a function returning a boolean value with the
\inlinecodevrs{#[pred]} decorator.
This explicitly names the conjuncts: e.g., \inlinecodevrs{valid} on line~\ref{line:unit:fn:valid}
states that the entry is valid if the present bit is set.

\myparagraph{Translation Function}
The \inlinecodevrs{translate()} function on line \ref{line:unit:fn:translate} defines how the output
address is computed.
The \inlinecodevrs{requires} clauses further establish the condition for the variant (i.e., the page
size bit must be one).

\myparagraph{Synthesis Target Functions}
Only the synthesis targets of \autoref{code:hlfn} are declared as \inlinecodevrs{synth fn}.
They define additional constraints on the synthesis target through pre-conditions, and
have the implicit post-conditions shown in \autoref{code:hlfn}.
Line~\ref{line:unit:fn:map} shows \osmap with the pre-conditions that establish constraints on the
alignment of the frame and the unit's supported mapping range, i.e., the mapping is
a large frame.

\subsection{Environment Specification}

Operating systems have different personalities, so the runtime environments (e.g., functions for
allocating and freeing memory, types, and means to access registers) vary.
For example, Linux allocates memory for the translation descriptors using
\inlinecodevrs{page_alloc}, whereas on capability-based systems (e.g., seL4~\cite{Klein:2009:SFV} or
Barrelfish~\cite{Baumann:2009:MNO}), applications perform allocations through a sequence of
capability operations, but the kernel still needs to format the translation descriptors correctly.
The description language allows developers to specify the runtime environment by declaring external
types and function signatures that the OS provides.
If changes can be applied only indirectly (e.g., through capability operations) the developer
provides the corresponding invocation to the \osfuns.

\section{Program Synthesis Methodology}
\label{sec:synthesis}

\system uses hierarchical software synthesis to automatically generate correct implementations for
the \osmap, \osprot and \osunmap functions (our \emph{synthesis} targets).
The resulting programs are correct, because \system verifies the synthesized programs against the
model and their specification, i.e., the post-conditions stated in \autoref{code:hlfn}.

We start by defining our synthesis problem, followed by an overview of the synthesis process.
We then describe the generation of candidate programs and the grammar that we use to construct them.
Finally, we present techniques to reduce the search space and handle caches, maintenance operations,
and barriers.

\subsection{Formulating the Synthesis Problem}
\label{sec:synth:problem}

\system solves the synthesis problem of finding a sequence of control interface invocations that
move the state machine from an arbitrary start state into an end state that has the desired
\hwmapping behavior, i.e., satisfies the post-condition of the synthesis target
(\autoref{code:hlfn}).

\begin{quote}
  \textbf{Synthesis Problem:}
Let $s_0$ represent an arbitrary starting state and $args$ represent the arguments to the higher-level
function. Further, let \texttt{assms} be the constraints on valid states, and \texttt{goal} be the post-condition of
the current synthesis target. Find a sequence of state transitions $op$ that satisfies the following
predicate.
\[
  \forall s_0, args ~::~ \texttt{assms}(s_0, args) \Rightarrow \texttt{goal}(\texttt{op}(s_0, args), args)
\]
\end{quote}

In other words, for any valid starting state and input arguments, applying the state transitions of
the candidate program results in a state that has the desired \hwmapping behavior.
The \texttt{assms} define assumptions on the initial state and the input arguments, e.g.,
the bit widths and alignments of addresses, and \texttt{goal} are the
post-conditions stated in \autoref{code:hlfn}.

\subsection{Synthesis Overview}
\label{sec:synth:overivew}

\begin{figure}
  \includegraphics{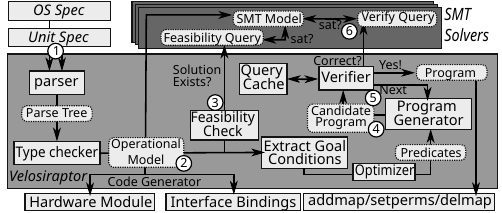}
  \caption{Synthesis Oveview}
  \label{fig:system}
\end{figure}

\autoref{fig:system} shows the overview of the synthesis process.
\circled{1} The input to the synthesis process is a specification of the \hwuint in the \system
specification language (\autoref{sec:language}).
\circled{2} From the specification, we create the operational model of the \hwuint including its
state and control interface.
\circled{3} We encode the operational model in smtlib2~\cite{smtlib2} using bit vector logic and
perform a feasibility check using the SMT solver.
\circled{4} The program generator starts constructing candidate programs, following a grammar
derived from the control interface.
\circled{5} We iterate through the candidate programs until we find one that satisfies the goal or
we exhaust the search space.
\circled{6} The verifier encodes the candidate program into an SMT query and uses an SMT solver
to check whether it satisfies the goal conditions derived from the post-conditions.
For efficiency, we apply various domain-specific optimizations that we describe in
\autoref{sec:synth:reduction}.

\subsection{Generating Candidate Programs}
\label{sec:synth:programs}

From the specifications of \osmap, \osprot, and \osunmap, \system automatically generates possible
implementations called \emph{candidate programs}.
With no constraints, this quickly leads to an explosion in the size of the search space.
We use a domain-specific property that lets us constrain the search space.
Recall, that the control interface of the \hwuint defines the software-visible API (i.e., the
interface fields as the function calls, and the bit slices are the argument values).
Thus, the control interface defines the grammar that generates the candidate programs.

\myparagraph{Grammar Description}
At the top level, a candidate program is a sequence of one or more atomic API invocations chosen
from the defined control interface fields.
Each API invocation is either a read or a write that will trigger the corresponding
\inlinecodevrs{ReadActions} or \inlinecodevrs{WriteActions}.
The API invocation consists of three parts: 1) initialize the argument value with zero, or read the
  current value from the control interface. 2) update the argument value by setting the values of
  the corresponding bit slices. 3) perform a read or write with the constructed argument value.
The values of the bit slices are set based on an expression that may involve the high-level function
parameters (e.g., address or size), binary operators, masks, or symbolic values.

\myparagraph{Program Generation}
Intuitively, the program generator starts by iterating over possible distinct combinations of API
invocations starting with the smallest possible candidate programs and then recursively expanding
them.
For example, assuming a \hwuint with two registers interface fields $A$ and $B$, the program
generator starts considering programs of the form \texttt{[A,B], [AA,AB,BA,BB]}, ...
For each of the API interactions, the program generator constructs all possible expressions that
update the argument value, initializations, and invocation type.
The number of possible programs grows factorially in the size of the interface.

\subsection{Search Space Reduction}
\label{sec:synth:reduction}

Simply generating all possible programs is impractical, so we introduce domain-specific search space
reduction techniques to make the synthesis process tractable.

\myparagraph{State Reduction Through Decomposition}
As outlined in \autoref{sec:methodology:buildingblocks}, we decompose complex \hwuint{}s into basic
building blocks and composition rules.
This reduces the state and the interface by splitting the \hwuint into smaller pieces.
For example, instead of considering an MMU with its multi-level page tables and registers in its
entirety, we can focus on a single page-table entry at a time.

\myparagraph{Program Structure}
The grammar that we use to construct candidate programs already constrains what programs we can
consider.
This not only reduces the search space but also enables the possibility of merging two programs,
a property that we leverage in the following technique.

\myparagraph{Divide-and-Conquer Synthesis Goal}
We leverage the definition of a successful \hwmap to split up the goal predicate, solve for each
subgoal independently, and merge the partial programs.
In \autoref{sec:methodology}, we defined a successful \hwmap operation as correctly calculating the
address and satisfying a conjunction of predicates.
The specification language (\autoref{code:language}) already follows this division.
We further rewrite all predicates into conjunctive normal form to enable further predicate
splitting.
We generate partial candidate programs and check whether they satisfy the subgoals.
Finally, we merge the partial candidate programs to obtain the candidate program and verify it
against the original goal.

\myparagraph{API Reduction}
The subgoal predicates may not use all bits of the state.
For example, the \inlinecodevrs{valid} predicate of \autoref{code:language} depends only on the
present bit.
Thus, instead of taking the entire control interface as the API, we reduce it to include only the
relevant parts by performing the following steps.
First, we create the subset of the state that is relevant to the current goal (i.e., parts of the
state that are referenced in the goal predicate).
Then we identify the parts of the interface that can change any of the relevant parts by looking at
the \inlinecodevrs{ReadActions} and \inlinecodevrs{WriteActions} blocks.
This is a form of back-projection of the bit slices of the interface onto the state.
This drastically reduces the grammar of the candidate programs to only 10-100 candidate programs per
subgoal.

\myparagraph{Expression Reduction}
The divide-and-conquer approach produces a set of subgoals that may have simple boolean expressions
such as $entry.present == 1$.
From the expression, we can extract various properties such as whether the returned value is a
boolean or address, whether any parameters of the synthesis target are accessed, the size of the bit
slice in bits, or whether the comparison is a constant expression.
This lets us reduce the possible expressions.
For example, if the expression does not refer to any of the arguments, we can exclude
those from the list of possible expressions.

\subsection{Tree-Based Divide-and-Conquer}
\label{sec:synth:opt:tree}
Instead of simply combining all partial programs of the conjuncts, we construct a balanced binary
tree structure.
The idea behind this is to filter wrong partial programs early to reduce the number of possible
combinations.
Consider an example with four conjuncts $A, B, C, D$ with $10$ candidate programs each, and
50\% of the possible programs for each subgoal satisfy the subgoal.
Combining all of them in one step ($A\wedge B\wedge C\wedge D$) produces $4\times 10 + 5^4 = 665$
programs to check.
However, with the tree structure $(A\wedge B)\wedge (C\wedge D)$, the number of programs drops to
$2\times (4\times 10 + 5^2) = 94$ for the leaf level, plus another $13^2=169$ candidate programs on
the middle level (with 50\% satisfiability).
This results in only $263$ programs to check compared to the $665$ programs that arise from brute
force search

\subsection{Caches and Barriers}
\label{sec:synth:barriers}

So far, we have assumed that interface interactions immediately update the state and become visible
to the \hwmapping function.
Unfortunately, this is not always true.
Stores may end up in write-buffers, caches, or even get reordered delaying the effect of the
interface interaction.

\myparagraph{Multi-Phase Synthesis}
\system handles cached or delayed updates by extending the \hwuint model and the program generator
to express write buffers and caches.
However, instead of directly including cache operations and barriers in the synthesis process, we
pursue a multi-phase approach.
The first phase operates as explained before and produces a program that satisfies the goal under
the assumption that updates become visible immediately.
The second phase takes the output program as a starting point and starts inserting barriers and
cache operations.
The new candidate programs are then verified against the extended model.

\myparagraph{Modeling Cache Operations}
Instead of applying the updates directly to the state, the extended model maintains a list of
outstanding write operations on the interface fields.
Read operations bypass this list and still return the current interface value directly.
We add additional operations for cache maintenance and barriers to the set of possible API
interactions that make sure all previous API interactions have been successfully executed.

\begin{codecfloat}{label=code:generated, caption={\system-generated Code for Mapping a Frame in the simplified x86 Page Directory from \autoref{code:language}.}, float}
size_t PDirEntryFrame_do_map(PDirEntryFrame_t * unit, vaddr_t va, size_t sz, flags_t flgs, paddr_t pa) {
  // pre-condition check: (pa & 0x1fffff) == 0x0
  if (!(((pa & 0x1fffff) == 0x0))) { return 0x0; }
  // configuration sequence
  PDirEntryFrame_entry_t v = PDirEntryFrame_entry_new(0);
  v = PDirEntryFrame_entry_address_set(v,
        ((pa >> 0x15) & 0x7ffffff));
  v = PDirEntryFrame_entry_ps_set(v, 0x1);
  v = PDirEntryFrame_entry_present_set(v, 0x1);
  v = PDirEntryFrame_entry_rw_set(v, flgs.writable);
  // write the entry
  PDirEntryFrame_entry__wr(unit, v);
  return LARGE_PAGE_SIZE;
}
\end{codecfloat}

\subsection{Code Generation}
\label{sec:synth:code}

The code generator takes the \hwuint specification, the synthesized sequences of interface
interactions, the OS environment specification, and produces the corresponding C or Rust source code
that implements the \osfuns functions.
\autoref{code:generated} shows the generated code for setting up a frame mapping for the simplified
x86 page directory from \autoref{code:language}.
For each of the defined building blocks and composition rules, the code generator declares a type
(e.g., \inlinecodevrs{PDirEntryFrame_t}).
For each field in the control interface, we generate a type with functions to set and extract bit
slice values (e.g., \inlinecodevrs{PDirEntryFrame_entry_present_set}).
The synthesized program then prepares the argument value \inlinecodevrs{v} and then invokes the
control interface (e.g., \inlinecodevrs{PDirEntryFrame_entry__wr}).
From the instantiated composition rules, the code generator emits the corresponding code following a
template.
Finally, the output is adapted to the OS environment.
The code generator finds the functions in the OS environment using the corresponding type signature.

The generated code follows a similar pattern to that found in Linux, seKVM~\cite{Li:2021:FVMP}, seL4~\cite{Klein:2009:SFV},
and Barrelfish~\cite{Baumann:2009:MNO}.
For example, \system generates functions that walk the page tables, allocate and free memory for page tables,
and write the corresponding page table entries as shown in \autoref{code:generated}.
Similarly, \system-generated code also includes types for the different unit types as seL4 and Barrelfish
do with its capability types or Linux and seKVM with their \inlinecodevrs{pmd_t}, \inlinecodevrs{pud_t}, etc,
respectively.

\section{Implementation}
\label{sec:implementation}

We describe the details of the \system implementation starting with an overview of the toolchain
implementation and how we encode the SMT queries.
We then explain our query scheduling and hardware generation backends, and show how \system helps
developers write specifications.

\subsection{Toolchain Overview}
\label{sec:impl:toolchain}

We implement \system in Rust and use the Z3 SMT solver~\cite{DeMoura:2008:Z3}.
The main components are the specification parsing, the intermediate representation of the \hwmapping
modules, the synthesizer and the program generator, and backends for code generation for OS code and
hardware components.
We run the Z3 SMT solver as multiple standalone processes to leverage parallelism.
\system sends smtlib2~\cite{smtlib2} formulas to the Z3 process and reads the results over a pipe.

The toolchain currently supports generating OS code in C and Rust and generating hardware components
for the Arm Fast Models Simulator.
We leave the generation of Verilog or other hardware description languages for future work.

\subsection{Encoding the \HwUnit Model in SMT}
\label{sec:impl:model}

\system converts the \hwuint specification into SMT formulas (smtlib2~\cite{smtlib2}) that can be
evaluated by an SMT solver.
The generated formulas can be divided into
1) a prelude that defines types,
2) a representation of the \hwuint model including its state and interface, and
3) queries that verify the candidate programs against the current goal.

\myparagraph{Type Encoding}
Each \hwuint has a specified input and output bit width it supports.
These bit widths define the maximum values for the \inlinecodevrs{vaddr}, \inlinecodevrs{paddr}, and
\inlinecodevrs{size} types of the specification language.
We declare a new typed term containing a bit vector of length 64 and a
boolean function constraining the supported range of the values.

\myparagraph{\HwUnit Encoding}
The encoding of a \hwuint model closely follows the language constructs (\autoref{code:language})
and the model outlined in \autoref{sec:methodology}.
We assert the model only once once for each solver instance.

We declare a new type containing a bit vector for each of the state and interface fields and
functions providing accessors to insert or extract the bit slices.
Then we combine the field definitions into data types for the interface and state, and combine the
interface and state datatypes into the model.
For the \inlinecodevrs{ReadActions} and \inlinecodevrs{WriteActions} we define a function with the
corresponding state updates.
We declare functions for predicates defined on the state, including the defined methods
in the \hwuint specification, pre-conditions, and other assumptions such as type constraints.

\myparagraph{Write Buffers}
To model caches and write buffers, we use Z3's internal list theory.
This adds a list of interface invocations with the corresponding arguments to the model state.
Executing the \inlinecodevrs{WriteActions} block just adds the operation to the list.
Cache maintenance operations or barriers will then trigger the execution of the corresponding
\inlinecodevrs{WriteActions} block of the operations in the list.

\subsection{Verifying Candidate Programs}
\label{sec:impl:programs}

To verify the candidate programs, we create a new assertion set and reuse the base model stated
above.
This enables checking the satisfiability of the current goal without re-instantiating the model
again.
\autoref{code:smt:query} shows this query.

First, we convert the implementation of the current candidate program into an SMT function
(Line~\ref{line:smt:prog}) that takes the initial model state plus arguments and produces a new
model state.
The body performs the sequence interface actions to update the model.
Secondly, we assert (Line~\ref{line:smt:assert}) that for all possible initial states and arguments,
the candidate program satisfies the goal condition for the end state.
We chose to use the \texttt{forall}-style instead of the negated \texttt{exists}, because it
allows us to directly obtain the value of symbolic variables.
Finally, we check whether the assertion holds (Line~\ref{line:smt:sat}).

\subsection{Use of SMT Output during Synthesis}
\label{sec:impl:smtoutput}

\begin{codez3float}{label=code:smt:query, caption={SMT query to verify candidate programs}, float}
(push)                       ; new assertion set
(declare-const svar!0 Num_t) ; define symbolic variable (*@ \label{line:smt:symvar} @*)
; current candidate program for map
(define-fun map ((st0 Model_t) (va VAddr_t) (sz Size_t) (*@ \label{line:smt:prog} @*)
                  (flgs Flags_t) (pa PAddr_t)) Model_t
  (let (st1 (Model.IFace.pte.set! st0 ZERO))
  (let (st2 (Model.IFace.pte.present.set! st1 symvar!0))
  (let (st3 (Model.IFace.pte.wr! st2)) st3))))
; assert post-condition of map
(assert (forall ((st!0 Model_t) (va VAddr_t) (pa PAddr_t)(*@\label{line:smt:assert} @*)
                 (sz Size_t)    (flgs Flags_t))
          (=> (map.assms st!0 va sz flgs pa)
               (translate.result (map st!0 va sz flgs pa)
                                  va sz flgs pa))))
(check-sat)           ; check satisfiability (*@\label{line:smt:sat} @*)
(get-value (svar!0))  ; obtain value of symbolic variable(*@\label{line:smt:val} @*)
(pop)                 ; drop the current assertion set
\end{codez3float}

\system uses the output of the SMT solver in three ways.
As described before, we use it to check whether a program satisfies the goal.
In addition, we use it for obtaining scalar values and detecting conflicting conditions.

\myparagraph{Obtaining Scalar Values}
During synthesis, we encounter expressions with unknown constants (e.g., \inlinecodevrs{addr << ?}).
We declare a symbolic variable for it (Line~\ref{line:smt:symvar}).
The SMT solver then tries to find a satisfying assignment to all variables including unknown constants.
If the SMT solver succeeds, we can extract the assigned value from the SMT output (Line \ref{line:smt:val})
and replace the unknown constant in the expression with the concrete value (e.g., \inlinecodevrs{addr << 12}).

\myparagraph{Conflicting Conditions}
As part of the synthesis process, we check whether the model itself is satisfiable, i.e., there
is a state that satisfies all mapping predicates.
We define a variable for the state, assert all mapping predicates, and then ask the SMT solver to
check satisfiability.
If it is not, we extract the \emph{unsat-core} from the SMT output, which shows which mapping predicates
are conflicting.

\subsection{Query Scheduling and Caching}
\label{sec:impl:querycache}

\system leverages caching and scheduling techniques to reduce the number of queries executed, thus
decreasing synthesis time.
We execute queries asynchronously to give more flexibility in scheduling them.
Query execution stops as soon as we find a program that satisfies the goal.

\myparagraph{Query Caching}
Before sending a query to the SMT solver, \system checks whether it already exists in the query
cache. If it does not, we add the query to the cache, mark the entry as pending, register the query,
execute it, and update the entry once the query is finished.
If the query is in the cache, we either return the cached result or register the query if the entry
was pending.
This avoids evaluating duplicate queries during the synthesis process.

\myparagraph{Prioritizing Queries}
We organize the execution of the SMT queries for the candidate programs as a tree (see
\autoref{sec:synth:reduction}).
The query scheduler assigns priorities to the SMT queries based on their position in the tree.
The root has the highest priority, and the leaves have the lowest priority.
The intuition behind this strategy is to prioritize the verification of more complete programs and
thus quickly bubble up candidate programs in the tree.
Ideally, this means that not all queries of the lower levels have to be executed.

\subsection{Hardware Generation}
\label{sec:impl:hwgen}

\system features a proof-of-concept hardware generation backend that produces a simulated \hwunit
that runs on the Arm Fast Models platform.
This platform allows for functional, but not cycle-accurate, simulation of a hardware platform.
In the Arm Fast Models, a platform is composed of multiple modules that are instantiated and their
ports (e.g., memory buses or interrupt lines) connected.
Each module is described as a LISA+~\cite{Arm:2020:LLFM} file, a domain-specific language built on
top of C++.
The Arm Fast Models toolchain then processes all LISA+ components and dependencies to build an
executable simulator binary.

\system generates a LISA+ file for the \hwunit
and a C++ implementation of the behavior for each unit defined in the specification file.
The \hwunit module has two upstream and downstream ports: two for the data plane (memory accesses),
one for the control interface (registers), and one for emitting memory accesses when the unit reads
in-memory descriptors.
The C++ implementation of each building block and composition rule is generated directly from its
specification; there is no synthesis required.

The generated LISA+ registers the \hwmapfn function of the top-level unit as the handler for
incoming memory accesses.
During the \hwmapping process, the code path follows the type hierarchy of the specification,
evaluating and calling the \hwmapfn function of the next unit, if permitted, or returning an error.
Ultimately, the output address is returned to the caller, and the incoming memory access is updated.
On error, the simulator raises an exception at the originating CPU core.

When describing a platform, the developer instantiates the \hwunit module and connects its ports to
the memory bus of a processor core and a memory module simulating RAM.
Multiple \hwunit modules can be connected serially, or different address ranges can be translated by
different \hwunit modules.
This enables the developer to set up complex memory hierarchies and translation schemes.

\section{Evaluation}
\label{sec:eval}

We evaluate \system using quantitative and qualitative scenarios to answer the following questions:

\begin{itemize}
    \item How quickly does \system synthesize code, and how well does synthesis scale?
      (\autoref{sec:eval:scaling})
    \item How much does each optimization improve runtime?
      (\autoref{sec:eval:optimizations})
    \item How well does the hardware software co-generation work? (\autoref{sec:eval:hardware})
    \item How easily can \system be integrated into operating systems? (\autoref{sec:eval:os})
    \item How does the performance of the generated code compare to hand-written code (\autoref{sec:eval:perf}).
\end{itemize}

\subsection{Evaluation Setup}

We run our evaluation on an Intel Xeon W-2275 processor with 14 cores with two hyperthreads each.
The host OS is Ubuntu 24.04 LTS with Linux kernel 6.5.0, and we use Z3 version 4.10.2 as the SMT
solver.
We set the number of workers equal to the number of cores.

\subsection{\system generates code quickly}
\label{sec:eval:scaling}

We evaluate \system synthesis and code generation performance showing that its runtime is
sufficiently small to be practically incorporated into kernel compilation.

\myparagraph{Methodology}
We measure the runtime of the synthesis and code generation for the translation hardware
descriptions shown in \autoref{tab:eval:synthesis}.
The units column shows the complexity of the description.
This has a direct correlation with the size of the grammar for the synthesized programs, larger
numbers indicate a bigger search space.
We also measure the compilation time of the Ubuntu Linux kernel v6.2.0 (default Ubuntu
configuration, no modules, using 28 hardware threads).

\begin{table}[t]
\begin{footnotesize}
\begin{center}
  \begin{tabular}{lr|rrr}
    \hline %
    \multicolumn{2}{c}{\textbf{Configurations}} & \multicolumn{3}{c}{\textbf{Results [ms]}} \\
    Name                 & \# Units             &  Programs   &   Time P50   &   Time P95   \\
    \hline %
    Simple Page Table    &          3U+ 2F+ 7S  &  9M+ 4P+ 3U &       151ms  &       163ms  \\
    x86\_32 Page Table   &          7U+ 5F+38S  & 39M+16P+ 9U &       561ms  &       582ms  \\
    x86\_64 Page Table   &         13U+ 8F+91S  & 72M+32P+18U &     1,137ms  &     1,171ms  \\
    Arm MPU              &          1U+ 7F+22S  &  9M+ 7P+ 3U &       750ms  &     1,129ms  \\
    Xeon Phi SMPT        &          2U+ 1F+ 3S  &  4M+ 3P+ 0U &        99ms  &       112ms  \\
    Simple Segment       &          1U+ 1F+ 2S  &  4M+ 0P+ 3U &        32ms  &        37ms  \\
    Variable Segment     &          1U+ 2F+ 5S  &  9M+ 4P+ 3U &       453ms  &       525ms  \\
    Medium Segment       &          1U+ 4F+18S  & 15M+ 5P+ 3U &       824ms  &     1,024ms  \\
    Assoc Segment        &          1U+ 4F+ 6S  & 14M+ 4P+ 3U &     1,058ms  &     4,475ms  \\
    x86 Segmentation     &          1U+ 2F+16S  & 17M+ 5P+ 3U &    12,445ms  &    12,683ms  \\
    R4700 TLB            &          3U+ 8F+13S  & 17M+ 0P+ 7U &     2,249ms  &     2,433ms  \\
    \hline %
    Linux Ubuntu         &                  --  & --          &   153,831ms  &   154,352ms  \\
    \hline %
  \end{tabular}

  \end{center}
  \end{footnotesize}
  \smallskip
  \caption{Synthesis Times and Complexities. The unit column shows the size of the specification
  in terms of basic building blocks and composition rules (U), the total size of the interface (F),
  and the number of bit slices (S). The program column shows the length of the synthesized programs
  for the \osmap (M), \osprot (P), and \osunmap (U) functions.}
  \label{tab:eval:synthesis}
\end{table}

\myparagraph{Results}
\autoref{tab:eval:synthesis} shows the median and 95th percentile runtimes over 100 runs.
Some of the hardware (e.g., Xeon Phi SMPT) does not support unmapping and/or access control so no
program is generated.
The measurements show that \system runtime is negligible relative to the kernel compilation.
For example, code synthesis for the x86\_64 page tables takes about one second, whereas the entire
kernel build takes more than 154 seconds.
We see that x86 segmentation has a significantly higher runtime compared to the x86\_64 page tables,
despite similar search space size (\autoref{sec:eval:optimizations}).
The page tables are very amenable to our divide-and-conquer approach resulting in many small subproblems,
whereas the x86 segmentation has a complex expression for the output address and limit calculations
that cannot be further split.

We see higher variability in the runtime for specifications that involve variable \hwmapping ranges
(i.e., Arm MPU, Variable, Medium and Assoc Segments).
Upon further examination, we observe a bimodal runtime distribution for these specifications.
Investigation of the generated SMT queries reveals a nested forall quantifier involving symbolic
variables.
Solving these queries may take a few milliseconds or can take several seconds to solve,
contributing to runtime variability and comprising almost half of the total runtime.

\myparagraph{Discussion}
\system provides a viable approach for automatically generating low-level
OS code without significantly affecting kernel build time.

\subsection{Ablation Study of Optimizations}
\label{sec:eval:optimizations}

We now evaluate how effectively each optimization reduces the search space size.

\myparagraph{Methodology}
We run \system on the descriptions shown in \autoref{tab:eval:synthesis} and selectively turn on the
optimizations outlined in \autoref{sec:synth:reduction} and calculate the search space of possible
candidate programs.
Due to the search space size, we are unable to run synthesis without certain optimizations and
therefore do not measure synthesis time.

\myparagraph{Results}
\autoref{tab:eval:opt} shows the calculated search space sizes (in base-10 logarithm) for the
different, cumulative optimization levels.
Enabling interface and state reduction optimizations together has the most significant effect on the
search space size, because it effectively reduces the size of the grammar with which we generate
programs.
Moreover, optimizing the possible set of expressions further reduces the search space significantly.
The tree-based structure does not reduce the maximum number of candidate programs, but instead,
quickly eliminates programs that do not satisfy the goal.

\begin{table}[t]
  \begin{footnotesize}

  \begin{tabular}{lrrrrr}
    \hline %
    Spec                 & No Opts   & +IfaceOpt & +StateOpt & +ExprOpt  & +Tree     \\
    \hline %
    Simple Page Table    &       23  &       23  &        5  &        3  &        3  \\
    x86\_32 Page Table   &       37  &       37  &       18  &       10  &       10  \\
    x86\_64 Page Table   &       38  &       38  &       21  &       13  &       13  \\
    Arm MPU              &       38  &       38  &        7  &        5  &        5  \\
    Xeon Phi SMPT        &        7  &        7  &        2  &        1  &        1  \\
    Simple Segment       &        4  &        4  &        2  &        1  &        1  \\
    Variable Segment     &       33  &       33  &        7  &        5  &        5  \\
    Medium Segment       &       37  &       37  &       13  &       10  &       10  \\
    Assoc Segment        &       38  &       38  &       12  &        9  &        9  \\
    x86 Segmentation     &       38  &       38  &       18  &       13  &       13  \\
    R4700 TLB            &       38  &       37  &       11  &        9  &        9  \\
    \hline %
  \end{tabular}

  \end{footnotesize}
  \smallskip
  \caption{Calculated search space sizes (log10) with cumulative optimizations.}
  \label{tab:eval:opt}
\end{table}

\myparagraph{Discussion}
This evaluation shows that domain-specific optimization techniques reduce the search space
significantly, making the synthesis process tractable for modern machines.

\subsection{Hardware/Software Co-Design with \system}
\label{sec:eval:hardware}

We demonstrate \system's ability to generate translation hardware components and the corresponding
code that configures it from the same description, using a qualitative study targeting the Arm Fast
Models platform simulator.

\myparagraph{Methodology}
We generate an emulated hardware component for the Arm Fast Models and instantiate it in the
platform description file by placing it between the main system bus and a DRAM region.
Memory accesses to a distinct, contiguous memory address range on the system bus will be forwarded
to the generated translation hardware component.
The hardware component \hwmap{}s or rejects the memory accesses based on the behavioral description
and its current state.
In case of a translation fault, a synchronous exception is raised to enable integration with the OS
as with MMU-based translation faults.
\system generates the hardware component and the corresponding OS code.
We boot our own simple, firmware-like image containing the \system-generated code
on the simulated hardware platform.
We then perform a sequence of loads and stores to the region translated by the generated hardware
component.
We reconfigure the memory hardware and repeat the test while checking whether we observe the expected behavior.

\myparagraph{Results and Discussion}
All cases booted successfully and we observed the expected behavior.
\system is capable of generating both a simulated hardware component and OS code from the same
description.
This is a promising result as it enables further research in \hwmapping hardware and may reduce the
effort of FPGA developers to use \hwmapping hardware in their FPGA-accelerated applications.

\subsection{\system can be used in a real OS}
\label{sec:eval:os}

We demonstrate that \system can be used in an existing OS code base by integrating
the generated code into the Barrelfish OS~\cite{Baumann:2009:MNO}, including a
monolithic setting.

\myparagraph{Methodology}
We take the x86\_64 page table specification and generate code for three OS environments:

\noindent\emph{1) Monolithic}: The code can directly access the registers and in-memory data
        structures that define the translation behavior. It also allocates and frees memory for the
        page tables.

\noindent\emph{2) uKernel supervisor}: The generated code can also directly access the registers and
        in-memory data structures, but it cannot allocate memory for the page tables.

\noindent\emph{3) uKernel userspace}: The generated code cannot access the registers or in-memory
        data structures directly, instead, invoking operations through system calls. It
        performs the required memory allocations for the translation descriptors.

We chose Barrelfish, a capability-based uKernel, as capability operations provide a nice separation
of tasks for managing page tables.
We add a \inlinecodevrs{mmap}-like system call to the kernel to test the monolithic
scenario.
For all three cases, we replace the existing code with the generated code.
We boot the system on x86\_64 Qemu with KVM acceleration such that the hardware performs the page
table walk where the guest page tables are set up by \system-generated code.
We then perform a sequence of memory mappings and accesses to the mapped regions to test whether
the memory hardware has been configured correctly.

\myparagraph{Results and Discussion}
We observed that the generated code correctly programs the translation hardware, the system booted,
and ran the tests successfully.
The OS environment specification can adapt the code generation of \system to handle a diverse set
of environments.

\subsection{Performance of Generated Code}
\label{sec:eval:perf}

We evaluate the performance of the generated code by comparing the runtime performance against
hand-written code of other operating systems.

\myparagraph{Methodology}
We run the evaluation entirely in userspace on Linux to avoid the overheads of system calls and rule
out possible differences in the OS environment.
The page-table structures are allocated on the heap.
We extract the hand-written x86\_64 page table code from the Barrelfish OS and the Linux kernel v6.8
and compare it against the \system-generated code.
We measure the latency of the \osfuns over 200 iterations.

\begin{table}
  \begin{footnotesize}
  \begin{tabular}{ccrrrrrr}
    \hline %
    \th{          }& \th{Operation } & \span{\th{  Map  }} & \span{\th{Protect}} & \span{\th{ Unmap }} \\
    \th{Structure }& \th{Code      } & \th{P50} & \th{P95} & \th{P50} & \th{P95} & \th{P50} & \th{P95} \\
    \hline %
    x86\_64        & Linux           &     13ns &     13ns &     13ns &     13ns &     12ns &     12ns \\
    x86\_64        & \system         &     13ns &     13ns &     12ns &     12ns &     11ns &     12ns \\
    \hline %
    PTable         & Barrelfish      &      8ns &      8ns &      8ns &      9ns &      7ns &      8ns \\
    PTable         & \system         &      8ns &      8ns &      8ns &      8ns &      8ns &      8ns \\
    \hline %
  \end{tabular}

  \end{footnotesize}
  \smallskip
  \caption{Performance comparison of \system-generated code with hand-written code from Linux and Barrelfish.}
  \label{tab:eval:perf}
\end{table}

\myparagraph{Results and Discussion}
\autoref{tab:eval:perf} shows the median and 95th percentile latencies of the three functions.
The x86\_64 structure sets up the full four-level page table structure in memory, whereas PTable
only configures the leaf level of the page table.
We observe that there is no significant difference between \system and the hand-written code.

\subsection{Discussion}

We demonstrated that \system produces performant code that configures the memory hardware in a wide
range of settings
Potential barriers to integrating the \system-generated code into a larger OS are replacing existing,
possibly tightly integrated data structures, e.g., Linux's paging data types being passed around
in the kernel instead of using high-level functions.
The OS may make use of non-present translation table entries to store information on
swapped-out pages, or during NUMA-balancing operations.
The high-level functions generated by \system do not expose functionality for these use cases, but
the enumeration construct could be used to express OS-specific use cases (e.g., swap entries). %

\section{Related Work}
\label{sec:relatedwork}

\myparagraph{MMU Configurations}
Achermann~\etal~\cite{Achermann:2021:GCIP} present a toolchain for automatically generating a
static initial page table for x86 and Arm-based systems based on a formal description of the memory
topology~\cite{Achermann:2017:FMA, Achermann:2018:PAR}.
Their work focused on calculating the right address mappings to install into the page table.
Fiedler~\etal~\cite{Fiedler:2023:SOS, Fiedler:2023:PHDF} define trust boundaries based on the
accessibility of control registers and translation tables.
D\"{o}rr~\etal~\cite{Dorr:2021:MCA} focus on access permissions and the generation of static
configurations.
In contrast, \system provides functions to dynamically change the configuration of memory hardware
at runtime and supports memory hardware besides translation tables.
Memory hardware management can also be split between user-space and kernel.
For example, Barrelfish~\cite{Baumann:2009:MNO}, seL4~\cite{Klein:2009:SFV} and
mmapx~\cite{Achermann:2021:MUM}, allow userspace to safely manage translation tables by
separating the construction from writing the actual bit patterns into the translation tables.
\system can handle generating the code for both the system call or capability invocations and the
kernel's updating of translation table entries (\autoref{sec:eval:os}).

\myparagraph{Device Driver Synthesis}
Termite~\cite{Ryzhyk:2009:ADD, Ryzhyk:2014:UDD} synthesizes device drivers using a game-theoretic
approach.
It expresses what happens when the OS interacts with device registers (e.g., the character sent
event), whereas \system focuses on the resulting \hwmapping, independent of the register
interactions.
Termite has no support for in-memory data structures and cannot generate corresponding hardware.
GhostWriter~\cite{Wang:2023:Ghostwriter} uses behavior trees to synthesize device drivers
by specifying the corresponding device behavior but it neither generates hardware
nor expresses \hwmapping behavior.

\myparagraph{Porting}
Porting an OS to new platforms requires significant effort~\cite{Rashid:1987:MIVA}.
Chen \etal~\cite{Chen:2014:DDG} show a signification reduction in developer time by generating
70\% of device driver code utilizing domain-specific aspects such as device classes and features.
\system follows a similar approach by synthesizing the memory management code.
Hu~\etal~\cite{Hu:2023:TPOS} present a system for automated porting of an OS to new
platforms using a machine-independent OS specification and a machine-dependent language describing
the instruction set.
Their work targets assembly code resulting in a huge search space, whereas \system operates on a
higher-level abstraction, i.e., API requests with a more manageable search space.

\myparagraph{Hardware Description}
Describing the hardware interface using a specification language allows for the generation of OS
bindings.
For example, the Mackerel language~\cite{mackerel} and Ghostwriter's register definition
file~\cite{Wang:2023:Ghostwriter} are domain-specific languages for describing registers and data
structures used by devices.
Neither includes the hardware behavior so cannot generate the hardware itself.
The Arm architecture is specified in ASL~\cite{Reid:2016:AASL} that serves as the
basis for generating the documentation, instruction interpreters~\cite{Reid:2020:UAAI},
simulators for processors, and test suites for chip designers~\cite{Reid:2016:TSAS}.

\myparagraph{MMU Virtualization}
Several works virtualize the MMU to provide isolation of system components and to enforce security
policies through intercepting and restricting access to the MMU configuration to validate configuration
updates.
Para-virtualization used in Xen~\cite{Barham:2003:Xen} and Linux's PV-Ops interface~\cite{linux:pvops}
intercepts updates to the page tables.
SecPod~\cite{Wang:2015:SecPod} isolates page-table management through paging delegation
and execution trapping to enforce auditing the kernel's page-table modifications by leveraging
the PV-Ops interface.
SecVisor~\cite{Seshadri:2007:SecVisor} virtualizes the MMU by controlling all modifications to the
page tables using nested paging and enforces policies on the execute permissions.
Hypervision~\cite{Azab:2014:Hypervision} validates page table updates in the secure monitor.
Nested kernel~\cite{Dautenhan:2015:NestedKernel} separates the kernel into a trusted, nested kernel
that has control over the MMU, and an untrusted outer kernel that has no direct control over the MMU.
Similarly, Nexen~\cite{Shi:2017:DeconstructingXen} uses the nested kernel architecture to introduce
a secure monitor into Xen by enforcing read-only page tables.
SVA~\cite{Criswell:2007:SVA} prevents the OS from mapping frames used by the ghost memory partition
or SVA VM internal memory as ordinary memory.
Inktag~\cite{Hofmann:2013:InkTag} divides physical memory into secure and non-secure frames
where secure frames can only be mapped by the trusted extended page table in a nested paging setup.
Overshadow~\cite{Chen:2008:Overshadow} protects applications from the OS by leveraging
context-dependent one-to-many mappings through the use of multiple shadow page tables.
Mondrix~\cite{Witchel:2005:Mondrix} overlays the address space with protection domains
assigning permissions to specific memory regions. %

Enforcing the security policies is orthogonal to \system.
Those systems could use \system-generated code in the ``bottom half'' of the virtualized MMU, i.e.,
the trusted component that directly interacts with the physical MMU.
Moreover, by tailoring the generated code with the OS specification, \system can provide
the corresponding para-virtualized interface that intercepts page table updates and invokes the
trusted component instead -- similar to the approach taken with Barrelfish in \autoref{sec:eval:os}.

\myparagraph{Hardware Mechanisms}
Several systems use hardware extensions to enforce isolation and protection.
Hardware-assisted kernel compartmentalization (HAKCs)~\cite{McKee:2022:HAKC} partitions the code and
data of a monolithic kernel and defines access policy for each partition using Arm's
PAC~\cite{arm:pac} and MTE~\cite{arm:mte} extensions.
xMP~\cite{Proskurin:2020:XMP} uses Intel's VMFunc~\cite{intel:sdm} to quickly switch the extended
page tables to accelerate protection domain switching.
vTZ~\cite{Hua:2017:VTZ} virtualizes Arm's secure world before hardware support has been
introduced~\cite{Arm:SecureVM} by leveraging the secure/non-secure modes and restricting second-stage
translation configuration to the secure world.
HDFI~\cite{Song::2016:HDFI} and CHERI~\cite{Watson:2015:CHERI} use tag-based memory protection
to perform an additional, hardware-accelerated permission check on every load/store instruction.
ACES~\cite{Clements:2018:ACES} uses a compiler-based approach to compartmentalize and isolate
components through binary instrumentation and uses memory protection units to enforce access
control.
SKEE~\cite{Azab:2016:SKEE} creates an isolated environment by disallowing the kernel from managing
its own translation tables, and forcing updates through a well-defined switch gate.
EC~\cite{Khan:2023:EC} uses a $\mu$Kernel-design to isolate components using a memory protection
unit and hardware watchpoints to monitor its configuration.
Kenali~\cite{Song:2016:EKSI} maintains a shadow address space with different access permissions
and enforces MMU integrity by restricting access to the MMU-related data and registers.

Leveraging hardware features such as Intel's VMFfunc, memory tagging or watchpoints, or maintaining
multiple page tables and shadow mappings is orthogonal to \system.
However, the \system-generated code could be used to set up the MMU-based address
translation, MPU-based memory protection, or create fat-pointers (e.g., CHERI ~\cite{Watson:2015:CHERI}).
We leave encryption-based approaches, e.g., pointer signatures, as future work.

\myparagraph{Formal Modeling and Verification}
KCoFI~\cite{Criswell:2014:KCoFI} provides an interface to safely manage the page tables for an
SVA VM~\cite{Criswell:2007:SVA} by assigning types to pages and then using this type information
when constructing page tables (similar to the typed capabilities in seL4~\cite{Klein:2009:SFV} and Barrelfish~\cite{Baumann:2009:MNO}).
\system defines a type per page table node to ensure correct construction.
Moreover, through OS specification the generated code can be tailored to API functions
(e.g., \texttt{sva.update.lN.mapping}) or the corresponding handler code in the kernel.
seKVM~\cite{Li:2021:FVMP} provides a model and a verified implementation for the ARMv8 virtual memory
system, including a proof that the resulting data structure is indeed a tree -- a property that
follows by construction from \system's specification language.
Moreover, seKVM has layers for allocation, walking, and constructing the page table, similar to the
generated functions of \system.
However, seKVM targets the ARMv8 architecture whereas \system allows generating
the corresponding functions for any given specification.
Li~\etal~\cite{Li:2022:DVAC} verified a concurrent implementation of multi-level realm translation
tables for the Arm architecture using their VIA framework.
The proof is tied to the Arm architecture and leverages the CertiKOS~\cite{Gu:2016:Certikos}
methodology.
CertiKOS itself uses a two-level page map that follows the x86\_32 page table structure.
\system could be used to generate code for other page tables and the corresponding proofs in the
VIA framework.
HyperKernel~\cite{Nelson:2017:Hyperkernel} ensures ownership properties of the root-level page table
to ensure safety and security guarantees.
With \system, porting Hyperkernel to other architectures could be simplified by generating the
code that configures the page tables.
Dai~\etal~\cite{Dai:2024:RustPT} verified the second stage translation setup of HyperEnclave,
focusing on a framework to verify the correctness using a deep embedding of Rust's MIR in Coq.
Brun~\etal~\cite{Brun:2023:BIOS} and Lattuada~\etal~\cite{Lattuada:2024:VPF} verify the correctness
of a Rust implementation of the x86\_64 page table in Verus with focus on hardware concurrency,
an aspect which we plan to integrate in the future.

\section{Conclusion}
\label{sec:conclusion}

Correctly configuring the memory hardware of a platform is of paramount importance for the system
integrity.
Hardware is becoming more diverse and complex, and OS developers must adapt their code to the
hardware they run on -- a manual process that takes time and requires developers to deal with
subtle, low-level hardware details.

\system is a methodology and system that uses software synthesis to automatically and efficiently
generate correct implementations for \osfuns to change the \hwmapping behavior of memory hardware.
\system lets developers generate the corresponding memory hardware itself, enabling
research into new \hwmapping schemes that precisely have the features the developer wants.
\begin{acks}
This work was supported by the Natural Sciences and Engineering Research Council of Canada (NSERC).
\end{acks}

\clearpage
\appendix
\section{Artifact}

\subsection{Abstract}

The artifact consists of
\begin{itemize}
  \item the \system toolchain,
  \item a set of \hwmapping specifications,
  \item a set of benchmarks / tests to reproduce the results in the paper, and
  \item and a modified version of the Barrelfish OS that uses the \system-generated code.
\end{itemize}
The following contains dependencies and instructions to reproduce the results presented in the
evaluation \autoref{sec:eval} of the paper.

\subsection{Artifact check-list (meta-information)}

{\small
\begin{itemize}
  \item {\bf Algorithm: Yes. }
  \item {\bf Compilation: Rust 1.79 and GCC 13.2.0 }
  \item {\bf Binary: requires Z3 SMT Solver, Arm FastModels, and Qemu}
  \item {\bf Run-time environment: Linux (Ubuntu 24.04)}
  \item {\bf Hardware: x86\_64 machine with a minimum of 8 cores and 32GB of RAM, either baremetal or a virtual machine}
  \item {\bf Run-time state: clean OS installation.}
  \item {\bf Execution: cargo / bash scripts.}
  \item {\bf Metrics: quantitative: performance (latency). Qualitative: functionality.}
  \item {\bf Output: Latex tables \ref{tab:eval:synthesis}, \ref{tab:eval:opt}, and \ref{tab:eval:perf}, code and OS logs. }
  \item {\bf Experiments: a total of five experiments each corresponding to a subsection in \autoref{sec:eval}.}
  \item {\bf How much disk space required?: 10GB.}
  \item {\bf How much time is needed to prepare workflow?: One hour. }
  \item {\bf How much time is needed to complete experiments?: Six hours. }
  \item {\bf Publicly available?: Yes, \url{https://github.com/ubc-systopia/velosiraptor-asplos25-artifact} }
  \item {\bf Code licenses?: MIT.}
  \item {\bf Data licenses?: None.}
  \item {\bf Workflow automation framework used?: No.}
  \item {\bf Archived (provide DOI)?: \href{https://doi.org/10.5281/zenodo.14752297}{10.5281/zenodo.14752297} or on \href{https://github.com/ubc-systopia/velosiraptor-asplos25-artifact}{GitHub repository}.}
\end{itemize}
}

\subsection{Description}

\subsubsection{How to access}

Clone the git repository at \url{https://github.com/ubc-systopia/velosiraptor-asplos25-artifact}
and follow the instructions in the README.md file.
Alternatively, download the artifact from \href{https://doi.org/10.5281/zenodo.14752297}{DOI: 10.5281/zenodo.14752297}
Ensure to initialize the submodules recursively.
The respository contains all the code and data sets, and instructions to install dependencies that are
required to reproduce the results on the machine.

\subsubsection{Hardware dependencies}

The artifact requires one x86\_64 machine (either baremetal or a virtual machine) with at least
eight cores and 32GB of RAM.
Smaller machines may also work, but can lead to longer runtimes or out-of-memory conditions.
Arm-based machines are not tested and only support a subset of the experiments and are thus
not recommended.
If a virtual machine is used, please ensure an x86\_64 host and appropriate configuration of hardware
acceleration (e.g., KVM) when running the performance benchmarks.

\subsubsection{Software dependencies}

The artifact is tested on x86\_64 Ubuntu 24.04 server -- other operating systems, e.g., other
Linux distributions and Windows Subsystem for Linux, may work too, but are not tested.
The README.md file provides a detailed list of software packages and other software dependencies
that can be installed using Ubuntu's package manager.
Additional dependencies are: Rust 1.79, Z3 4.10.2, Docker, and Arm Fast Models 11.15.

\subsection{Installation and Basic Test}

Follow the instructions in the \emph{Preparation} section of the README.md file
in the repository to setup the build and run environment of the artifact.
Beyond the dependencies, there is no installation necessary and the artifact builds and
runs out of the repository.

\textbf{Basic Test:} To test the artifact run the smoke test from section 6.2 and/or run the test
Section 6.3. This exercises the synthesis workflow, SMT solver, optimizations, and code compilation.

\subsection{Experiment workflow}

Most of the experiments are run through \verb|cargo| commands or bash scripts
as described in the README.md file of the artifact repository.
The workflow consists of two phases:
\begin{enumerate}
  \item running the respective commands to execute the experiment, and
  \item inspecting the output to verify and compare the results.
\end{enumerate}
The README.md contains detailed descriptions of the commands and the expected output.

\subsection{Evaluation and expected results}

The artifact is expected to
\begin{enumerate}
  \item closely reproduce Tables \ref{tab:eval:synthesis}, \ref{tab:eval:opt}, and \ref{tab:eval:perf}
        in the evaluation \autoref{sec:eval} of the paper (quantitative benchmarks), and
  \item emit similar output as described in the README.md file for the remaining experiments of
        Sections \ref{sec:eval:hardware} and \ref{sec:eval:os} in the paper (qualitative benchmarks).
\end{enumerate}

\subsection{Experiment customization}

Most experiments run within a few minutes in the current configuration.
For the longer-running synthesis experiment (\autoref{tab:eval:synthesis}), we provide a smoke
test that runs a smaller version of the experiment.
Moreover, experiments can also be customized by changing the set of specifications that
are evaluated, or the number of iterations executed.
This requires modifying the respective benchmarking code in the repository.

\subsection{Notes}

\paragraph{Arm Fast Models Licenses}
The emulated hardware platforms require the Arm Fast Models to build and run the simulators.
The Arm Fast Models require licenses.
To download the Arm Fast Models and and obtaining the corresponding licenses, visit the
Arm website: \url{https://www.arm.com/products/development-tools/simulation/fast-models}.
The artifact assumes that the user has access to
\begin{enumerate}
  \item the Arm Fast Models and its third-party IP,
  \item license for generating the simulation platform,
  \item license to run the base platform, and
  \item license to run the Arm Cortex A57 processor IP.
\end{enumerate}

\bibliographystyle{ACM-Reference-Format}
\balance
\bibliography{content/references}

@inproceedings{Achermann:2017:FMA,
  author     = {Reto Achermann and Lukas Humbel and David Cock and Timothy Roscoe},
  bibsource  = {dblp computer science bibliography, https://dblp.org},
  booktitle  = {Proceedings 2nd Workshop on Models for Formal Analysis of Real Systems, MARS at ETAPS 2017, Uppsala, Sweden, 29th April 2017.},
  conference = {MARS'17},
  date       = {2017},
  doi        = {10.4204/EPTCS.244.4},
  pages      = {66--116},
  series     = {MARS'17},
  title      = {Formalizing Memory Accesses and Interrupts},
  url        = {https://doi.org/10.4204/EPTCS.244.4},
  year       = {2017}
}

@inproceedings{Achermann:2018:PAR,
  author     = {Reto Achermann and Lukas Humbel and David Cock and Timothy Roscoe},
  bibsource  = {dblp computer science bibliography, https://dblp.org},
  booktitle  = {Proceedings of the 9th International Conference on Interactive Theorem Proving, 2018, Held as Part of the Federated Logic Conference, FloC 2018},
  conference = {ITP'18},
  date       = {2018},
  doi        = {10.1007/978-3-319-94821-8_1},
  location   = {Oxford, UK},
  pages      = {1--19},
  series     = {ITP'18},
  title      = {{Physical Addressing on Real Hardware in Isabelle/HOL}},
  url        = {https://doi.org/10.1007/978-3-319-94821-8_1},
  year       = {2018}
}

@inproceedings{Achermann:2021:GCIP,
  address   = {New York, NY, USA},
  author    = {Achermann, Reto and Cock, David and Haecki, Roni and Hossle, Nora and Humbel, Lukas and Roscoe, Timothy and Schwyn, Daniel},
  booktitle = {Proceedings of the 11th Workshop on Programming Languages and Operating Systems},
  date      = {2021},
  doi       = {10.1145/3477113.3487270},
  isbn      = {9781450387071},
  location  = {Virtual Event, Germany},
  numpages  = {7},
  pages     = {69–75},
  publisher = {Association for Computing Machinery},
  series    = {PLOS '21},
  title     = {{Generating Correct Initial Page Tables from Formal Hardware Descriptions}},
  url       = {https://doi.org/10.1145/3477113.3487270},
  venue     = {PLOS '21},
  year      = {2021}
}

@inproceedings{Achermann:2021:MUM,
  address   = {New York, NY, USA},
  author    = {Achermann, Reto and Cock, David and Haecki, Roni and Hossle, Nora and Humbel, Lukas and Roscoe, Timothy and Schwyn, Daniel},
  booktitle = {Proceedings of the Workshop on Hot Topics in Operating Systems},
  date      = {2021},
  doi       = {10.1145/3458336.3465273},
  isbn      = {9781450384384},
  location  = {Ann Arbor, Michigan},
  numpages  = {8},
  pages     = {159–166},
  publisher = {Association for Computing Machinery},
  series    = {HotOS '21},
  title     = {mmapx: uniform memory protection in a heterogeneous world},
  url       = {https://doi.org/10.1145/3458336.3465273},
  venue     = {HotOS '21},
  year      = {2021}
}

@inproceedings{Alam:2017:DVMT,
  author    = {Alam, Hanna and Zhang, Tianhao and Erez, Mattan and Etsion, Yoav},
  title     = {Do-It-Yourself Virtual Memory Translation},
  year      = {2017},
  isbn      = {9781450348928},
  publisher = {Association for Computing Machinery},
  address   = {New York, NY, USA},
  url       = {https://doi.org/10.1145/3079856.3080209},
  doi       = {10.1145/3079856.3080209},
  booktitle = {Proceedings of the 44th Annual International Symposium on Computer Architecture},
  pages     = {457–468},
  numpages  = {12},
  location  = {Toronto, ON, Canada},
  series    = {ISCA '17}
}

@manual{Arm:2020:LLFM,
  title  = {LISA+ Language for Fast Models Reference Manual},
  author = {{Arm Ltd.}},
  year   = {2023},
  note   = {Version 1.0.},
  url    = {https://developer.arm.com/documentation/101092/0100}
}

@misc{Arm:2024:FastModels,
  author       = {{Arm Ltd.}},
  title        = {Fast Models: SoC Verification Without Hardware},
  howpublished = {Online. \url{https://www.arm.com/products/development-tools/simulation/fast-models}},
  year         = {2024}
}

@misc{Arm:SecureVM,
  author       = {Berenice Mann},
  title        = {New Secure world architecture in Armv8.4},
  howpublished = {Online. \url{https://community.arm.com/arm-community-blogs/b/architectures-and-processors-blog/posts/architecting-more-secure-world-with-isolation-and-virtualization}},
  year         = {2018}
}

@inproceedings{Azab:2014:Hypervision,
  author    = {Azab, Ahmed M. and Ning, Peng and Shah, Jitesh and Chen, Quan and Bhutkar, Rohan and Ganesh, Guruprasad and Ma, Jia and Shen, Wenbo},
  title     = {Hypervision Across Worlds: Real-time Kernel Protection from the ARM TrustZone Secure World},
  year      = {2014},
  isbn      = {9781450329576},
  publisher = {Association for Computing Machinery},
  address   = {New York, NY, USA},
  url       = {https://doi.org/10.1145/2660267.2660350},
  doi       = {10.1145/2660267.2660350},
  booktitle = {Proceedings of the 2014 ACM SIGSAC Conference on Computer and Communications Security},
  pages     = {90–102},
  numpages  = {13},
  location  = {Scottsdale, Arizona, USA},
  series    = {CCS '14}
}

@inproceedings{Azab:2016:SKEE,
  title     = {SKEE: A lightweight Secure Kernel-level Execution Environment for ARM},
  author    = {Ahmed M. Azab and Kirk Swidowski and Rohan Bhutkar and Jia Ma and Wenbo Shen and Ruowen Wang and Peng Ning},
  booktitle = {Network and Distributed System Security Symposium},
  publisher = {The Internet Society},
  series    = {NDSS'16},
  location  = {San Diego, California, USA},
  year      = {2016},
  doi       = {10.14722/ndss.2016.23009}
}

@inproceedings{Barham:2003:Xen,
  author    = {Barham, Paul and Dragovic, Boris and Fraser, Keir and Hand, Steven and Harris, Tim and Ho, Alex and Neugebauer, Rolf and Pratt, Ian and Warfield, Andrew},
  title     = {Xen and the art of virtualization},
  year      = {2003},
  isbn      = {1581137575},
  publisher = {Association for Computing Machinery},
  address   = {New York, NY, USA},
  url       = {https://doi.org/10.1145/945445.945462},
  doi       = {10.1145/945445.945462},
  booktitle = {Proceedings of the Nineteenth ACM Symposium on Operating Systems Principles},
  pages     = {164–177},
  numpages  = {14},
  location  = {Bolton Landing, NY, USA},
  series    = {SOSP '03}
}

@article{Basu:2023:EVM,
  author     = {Basu, Arkaprava and Gandhi, Jayneel and Chang, Jichuan and Hill, Mark D. and Swift, Michael M.},
  title      = {Efficient virtual memory for big memory servers},
  year       = {2013},
  issue_date = {June 2013},
  publisher  = {Association for Computing Machinery},
  address    = {New York, NY, USA},
  volume     = {41},
  number     = {3},
  issn       = {0163-5964},
  url        = {https://doi.org/10.1145/2508148.2485943},
  doi        = {10.1145/2508148.2485943},
  journal    = {SIGARCH Comput. Archit. News},
  month      = {jun},
  pages      = {237–248},
  numpages   = {12}
}

@inproceedings{Baumann:2009:MNO,
  author    = {Baumann, Andrew and Barham, Paul and Dagand, Pierre-Evariste and Harris, Tim and Isaacs, Rebecca and Peter, Simon and Roscoe, Timothy and Sch\"{u}pbach, Adrian and Singhania, Akhilesh},
  title     = {The multikernel: a new OS architecture for scalable multicore systems},
  year      = {2009},
  isbn      = {9781605587523},
  publisher = {Association for Computing Machinery},
  address   = {New York, NY, USA},
  url       = {https://doi.org/10.1145/1629575.1629579},
  doi       = {10.1145/1629575.1629579},
  booktitle = {Proceedings of the ACM SIGOPS 22nd Symposium on Operating Systems Principles},
  pages     = {29–44},
  numpages  = {16},
  location  = {Big Sky, Montana, USA},
  series    = {SOSP '09}
}

@inproceedings{Chen:2008:Overshadow,
  author    = {Chen, Xiaoxin and Garfinkel, Tal and Lewis, E. Christopher and Subrahmanyam, Pratap and Waldspurger, Carl A. and Boneh, Dan and Dwoskin, Jeffrey and Ports, Dan R.K.},
  title     = {Overshadow: a virtualization-based approach to retrofitting protection in commodity operating systems},
  year      = {2008},
  isbn      = {9781595939586},
  publisher = {Association for Computing Machinery},
  address   = {New York, NY, USA},
  url       = {https://doi.org/10.1145/1346281.1346284},
  doi       = {10.1145/1346281.1346284},
  booktitle = {Proceedings of the 13th International Conference on Architectural Support for Programming Languages and Operating Systems},
  pages     = {2–13},
  numpages  = {12},
  location  = {Seattle, WA, USA},
  series    = {ASPLOS XIII}
}

@inproceedings{Chen:2014:DDG,
  author    = {H. {Chen} and G. {Godet-Bar} and F. {Rousseau} and F. {Petrot}},
  booktitle = {2014 25nd IEEE International Symposium on Rapid System Prototyping},
  title     = {Device driver generation targeting multiple operating systems using a model-driven methodology},
  year      = {2014},
  volume    = {},
  number    = {},
  pages     = {30-36},
  doi       = {10.1109/RSP.2014.6966689}
}

@inproceedings{Clements:2018:ACES,
  author    = {Clements, Abraham A. and Almakhdhub, Naif Saleh and Bagchi, Saurabh and Payer, Mathias},
  title     = {ACES: automatic compartments for embedded systems},
  year      = {2018},
  isbn      = {9781931971461},
  publisher = {USENIX Association},
  address   = {USA},
  booktitle = {Proceedings of the 27th USENIX Conference on Security Symposium},
  pages     = {65–82},
  numpages  = {18},
  location  = {Baltimore, MD, USA},
  series    = {SEC'18}
}

@inproceedings{Criswell:2007:SVA,
  author    = {Criswell, John and Lenharth, Andrew and Dhurjati, Dinakar and Adve, Vikram},
  title     = {Secure virtual architecture: a safe execution environment for commodity operating systems},
  year      = {2007},
  isbn      = {9781595935915},
  publisher = {Association for Computing Machinery},
  address   = {New York, NY, USA},
  url       = {https://doi.org/10.1145/1294261.1294295},
  doi       = {10.1145/1294261.1294295},
  booktitle = {Proceedings of Twenty-First ACM SIGOPS Symposium on Operating Systems Principles},
  pages     = {351–366},
  numpages  = {16},
  location  = {Stevenson, Washington, USA},
  series    = {SOSP '07}
}

@inproceedings{Criswell:2014:KCoFI,
  author    = {Criswell, John and Dautenhahn, Nathan and Adve, Vikram},
  title     = {KCoFI: Complete Control-Flow Integrity for Commodity Operating System Kernels},
  year      = {2014},
  isbn      = {9781479946860},
  publisher = {IEEE Computer Society},
  address   = {USA},
  url       = {https://doi.org/10.1109/SP.2014.26},
  doi       = {10.1109/SP.2014.26},
  booktitle = {Proceedings of the 2014 IEEE Symposium on Security and Privacy},
  pages     = {292–307},
  numpages  = {16},
  series    = {SP '14}
}

@misc{CVE-2014-9888,
  author       = {{National Vulnerability Database}},
  howpublished = {Online.\url{https://nvd.nist.gov/vuln/detail/CVE-2014-9888}},
  title        = {{CVE-2014-9888}},
  year         = {2014}
}

@misc{CVE-2017-16994,
  author       = {{National Vulnerability Database}},
  howpublished = {Online. \url{https://nvd.nist.gov/vuln/detail/CVE-2017-16994}},
  title        = {{CVE-2017-16994}},
  year         = {2017}
}

@inproceedings{Dai:2024:RustPT,
  author    = {Dai, Zhenyang and Liu, Shuang and Sjoberg, Vilhelm and Li, Xupeng and Chen, Yu and Wang, Wenhao and Jia, Yuekai and Anderson, Sean Noble and Elbeheiry, Laila and Sondhi, Shubham and Zhang, Yu and Ni, Zhaozhong and Yan, Shoumeng and Gu, Ronghui and He, Zhengyu},
  title     = {Verifying Rust Implementation of Page Tables in a Software Enclave Hypervisor},
  year      = {2024},
  isbn      = {9798400703850},
  publisher = {Association for Computing Machinery},
  address   = {New York, NY, USA},
  url       = {https://doi.org/10.1145/3620665.3640398},
  doi       = {10.1145/3620665.3640398},
  booktitle = {Proceedings of the 29th ACM International Conference on Architectural Support for Programming Languages and Operating Systems, Volume 2},
  pages     = {1218–1232},
  numpages  = {15},
  location  = {La Jolla, CA, USA},
  series    = {ASPLOS '24}
}

@inproceedings{Dautenhan:2015:NestedKernel,
  author    = {Dautenhahn, Nathan and Kasampalis, Theodoros and Dietz, Will and Criswell, John and Adve, Vikram},
  title     = {Nested Kernel: An Operating System Architecture for Intra-Kernel Privilege Separation},
  year      = {2015},
  isbn      = {9781450328357},
  publisher = {Association for Computing Machinery},
  address   = {New York, NY, USA},
  url       = {https://doi.org/10.1145/2694344.2694386},
  doi       = {10.1145/2694344.2694386},
  booktitle = {Proceedings of the Twentieth International Conference on Architectural Support for Programming Languages and Operating Systems},
  pages     = {191–206},
  numpages  = {16},
  location  = {Istanbul, Turkey},
  series    = {ASPLOS '15}
}

@inproceedings{DeMoura:2008:Z3,
  author    = {De Moura, Leonardo and Bj\o{}rner, Nikolaj},
  title     = {Z3: an efficient SMT solver},
  year      = {2008},
  isbn      = {3540787992},
  publisher = {Springer-Verlag},
  address   = {Berlin, Heidelberg},
  booktitle = {Proceedings of the Theory and Practice of Software, 14th International Conference on Tools and Algorithms for the Construction and Analysis of Systems},
  pages     = {337–340},
  numpages  = {4},
  location  = {Budapest, Hungary},
  series    = {TACAS'08/ETAPS'08}
}

@article{Dorr:2021:MCA,
  author     = {D\"{o}rr, Tobias and Sandmann, Timo and Becker, J\"{u}rgen},
  title      = {Model-based configuration of access protection units for multicore processors in embedded systems},
  year       = {2021},
  issue_date = {Nov 2021},
  publisher  = {Elsevier Science Publishers B. V.},
  address    = {NLD},
  volume     = {87},
  number     = {C},
  issn       = {0141-9331},
  url        = {https://doi.org/10.1016/j.micpro.2021.104377},
  doi        = {10.1016/j.micpro.2021.104377},
  journal    = {Microprocess. Microsyst.},
  month      = {nov},
  numpages   = {17}
}

@inproceedings{Fiedler:2023:PHDF,
  author    = {Fiedler, Ben and Schwyn, Daniel and Gierczak-Galle, Constantin and Cock, David and Roscoe, Timothy},
  title     = {Putting out the hardware dumpster fire},
  year      = {2023},
  isbn      = {9798400701955},
  publisher = {Association for Computing Machinery},
  address   = {New York, NY, USA},
  url       = {https://doi.org/10.1145/3593856.3595903},
  doi       = {10.1145/3593856.3595903},
  booktitle = {Proceedings of the 19th Workshop on Hot Topics in Operating Systems},
  pages     = {46–52},
  numpages  = {7},
  location  = {Providence, RI, USA},
  series    = {HOTOS '23}
}

@inproceedings{Fiedler:2023:SOS,
  author    = {Fiedler, Ben and Meier, Roman and Schult, Jasmin and Schwyn, Daniel and Roscoe, Timothy},
  title     = {Specifying the de-facto OS of a production SoC},
  year      = {2023},
  isbn      = {9798400704116},
  publisher = {Association for Computing Machinery},
  address   = {New York, NY, USA},
  url       = {https://doi.org/10.1145/3625275.3625400},
  doi       = {10.1145/3625275.3625400},
  booktitle = {Proceedings of the 1st Workshop on Kernel Isolation, Safety and Verification},
  pages     = {18–25},
  numpages  = {8},
  location  = {Koblenz, Germany},
  series    = {KISV '23}
}

@inproceedings{Gosakan:2023:MPBT,
  author    = {Gosakan, Krishnan and Han, Jaehyun and Kuszmaul, William and Mubarek, Ibrahim N. and Mukherjee, Nirjhar and Sriram, Karthik and Tagliavini, Guido and West, Evan and Bender, Michael A. and Bhattacharjee, Abhishek and Conway, Alex and Farach-Colton, Martin and Gandhi, Jayneel and Johnson, Rob and Kannan, Sudarsun and Porter, Donald E.},
  title     = {Mosaic Pages: Big TLB Reach with Small Pages},
  year      = {2023},
  isbn      = {9781450399180},
  publisher = {Association for Computing Machinery},
  address   = {New York, NY, USA},
  url       = {https://doi.org/10.1145/3582016.3582021},
  doi       = {10.1145/3582016.3582021},
  booktitle = {Proceedings of the 28th ACM International Conference on Architectural Support for Programming Languages and Operating Systems, Volume 3},
  pages     = {433–448},
  numpages  = {16},
  location  = {Vancouver, BC, Canada},
  series    = {ASPLOS 2023}
}

@inproceedings{Gu:2016:Certikos,
  author    = {Ronghui Gu and Zhong Shao and Hao Chen and Xiongnan (Newman) Wu and Jieung Kim and Vilhelm Sj{\"o}berg and David Costanzo},
  title     = {{CertiKOS}: An Extensible Architecture for Building Certified Concurrent {OS} Kernels},
  booktitle = {12th USENIX Symposium on Operating Systems Design and Implementation (OSDI 16)},
  year      = {2016},
  isbn      = {978-1-931971-33-1},
  address   = {Savannah, GA},
  pages     = {653--669},
  url       = {https://www.usenix.org/conference/osdi16/technical-sessions/presentation/gu},
  publisher = {USENIX Association},
  month     = nov
}

@inproceedings{Hofmann:2013:InkTag,
  author    = {Hofmann, Owen S. and Kim, Sangman and Dunn, Alan M. and Lee, Michael Z. and Witchel, Emmett},
  title     = {InkTag: secure applications on an untrusted operating system},
  year      = {2013},
  isbn      = {9781450318709},
  publisher = {Association for Computing Machinery},
  address   = {New York, NY, USA},
  url       = {https://doi.org/10.1145/2451116.2451146},
  doi       = {10.1145/2451116.2451146},
  booktitle = {Proceedings of the Eighteenth International Conference on Architectural Support for Programming Languages and Operating Systems},
  pages     = {265–278},
  numpages  = {14},
  location  = {Houston, Texas, USA},
  series    = {ASPLOS '13}
}

@inproceedings{hotos,
author = {Achermann, Reto and Karimalis, Ilias and Seltzer, Margo},
title = {Why write address translation OS code yourself when you can synthesize it?},
year = {2023},
isbn = {9798400701955},
publisher = {Association for Computing Machinery},
address = {New York, NY, USA},
url = {https://doi.org/10.1145/3593856.3595895},
doi = {10.1145/3593856.3595895},
booktitle = {Proceedings of the 19th Workshop on Hot Topics in Operating Systems},
pages = {174–180},
numpages = {7},
location = {Providence, RI, USA},
series = {HOTOS '23}
}

@inproceedings{Lattuada:2024:VPF,
author = {Lattuada, Andrea and Hance, Travis and Bosamiya, Jay and Brun, Matthias and Cho, Chanhee and LeBlanc, Hayley and Srinivasan, Pranav and Achermann, Reto and Chajed, Tej and Hawblitzel, Chris and Howell, Jon and Lorch, Jacob R. and Padon, Oded and Parno, Bryan},
title = {Verus: A Practical Foundation for Systems Verification},
year = {2024},
isbn = {9798400712517},
publisher = {Association for Computing Machinery},
address = {New York, NY, USA},
url = {https://doi.org/10.1145/3694715.3695952},
doi = {10.1145/3694715.3695952},
booktitle = {Proceedings of the ACM SIGOPS 30th Symposium on Operating Systems Principles},
pages = {438–454},
numpages = {17},
location = {Austin, TX, USA},
series = {SOSP '24}
}

@inproceedings{Brun:2023:BIOS,
  author    = {Brun, Matthias and Achermann, Reto and Chajed, Tej and Howell, Jon  and Zellweger, Gerd   and Lattuada, Andrea  },
  title     = {Beyond isolation: OS verification as a foundation for correct applications},
  year      = {2023},
  isbn      = {979-8-4007-0195-5},
  publisher = {Association for Computing Machinery},
  url       = {https://doi.org/10.1145/3593856.3595899},
  doi       = {10.1145/3593856.3595899},
  booktitle = {Proceedings of the 19th Workshop on Hot Topics in Operating Systems},
  pages     = {158--165},
  location  = {Providence, Rhode Island, USA},
  series    = {HotOS'23},
  tags      = {workshop}
}

@article{Hu:2023:TPOS,
  author     = {Hu, Jingmei and Lu, Eric and Holland, David A. and Kawaguchi, Ming and Chong, Stephen and Seltzer, Margo},
  title      = {Towards Porting Operating Systems with Program Synthesis},
  year       = {2023},
  issue_date = {March 2023},
  publisher  = {Association for Computing Machinery},
  address    = {New York, NY, USA},
  volume     = {45},
  number     = {1},
  issn       = {0164-0925},
  url        = {https://doi.org/10.1145/3563943},
  doi        = {10.1145/3563943},
  journal    = {ACM Trans. Program. Lang. Syst.},
  month      = {mar},
  articleno  = {2},
  numpages   = {70}
}

@inproceedings{Hua:2017:VTZ,
  author    = {Zhichao Hua and Jinyu Gu and Yubin Xia and Haibo Chen and Binyu Zang and Haibing Guan},
  title     = {{vTZ}: Virtualizing {ARM} {TrustZone}},
  booktitle = {26th USENIX Security Symposium (USENIX Security 17)},
  year      = {2017},
  isbn      = {978-1-931971-40-9},
  address   = {Vancouver, BC},
  pages     = {541--556},
  url       = {https://www.usenix.org/conference/usenixsecurity17/technical-sessions/presentation/hua},
  publisher = {USENIX Association},
  month     = aug
}

@inproceedings{Huang:2016:ESL,
  author    = {Huang, Jian and Qureshi, Moinuddin K. and Schwan, Karsten},
  title     = {{An Evolutionary Study of Linux Memory Management for Fun and
               Profit}},
  booktitle = {Proceedings of the 2016 USENIX Conference on Usenix Annual
               Technical Conference},
  series    = {USENIX ATC '16},
  year      = {2016},
  isbn      = {978-1-931971-30-0},
  location  = {Denver, CO, USA},
  pages     = {465--478},
  numpages  = {14},
  url       = {http://dl.acm.org/citation.cfm?id=3026959.3027002},
  acmid     = {3027002},
  publisher = {USENIX Association},
  address   = {Berkeley, CA, USA}
}

@manual{intel:sdm,
  title  = {{Intel 64 and IA-32 Architectures Software Developer's Manual}},
  author = {{Intel Corp.}},
  note   = {December 2023},
  url    = {https://www.intel.com/content/www/us/en/developer/articles/technical/intel-sdm.html},
  year   = {2023}
}

@manual{intel:xeonphi,
  title  = {Intel Xeon Phi Coprocessor Developer's Quick Start Guide},
  author = {{Intel Corp.}},
  note   = {Version 1.8},
  year   = {2013},
  url    = {https://www.intel.com/content/dam/develop/external/us/en/documents/intel-xeon-phi-coprocessor-quick-start-developers-guide.pdf}
}

@inproceedings{Khan:2023:EC,
  author    = {Khan, Arslan and Xu, Dongyan and Tian, Dave Jing},
  booktitle = {2023 IEEE Symposium on Security and Privacy (SP)},
  title     = {EC: Embedded Systems Compartmentalization via Intra-Kernel Isolation},
  year      = {2023},
  volume    = {},
  number    = {},
  pages     = {2990-3007},
  doi       = {10.1109/SP46215.2023.10179285}
}

@inproceedings{Klein:2009:SFV,
  author    = {Klein, Gerwin and Elphinstone, Kevin and Heiser, Gernot and Andronick, June and Cock, David and Derrin, Philip and Elkaduwe, Dhammika and Engelhardt, Kai and Kolanski, Rafal and Norrish, Michael and Sewell, Thomas and Tuch, Harvey and Winwood, Simon},
  title     = {seL4: formal verification of an OS kernel},
  year      = {2009},
  isbn      = {9781605587523},
  publisher = {Association for Computing Machinery},
  address   = {New York, NY, USA},
  url       = {https://doi.org/10.1145/1629575.1629596},
  doi       = {10.1145/1629575.1629596},
  booktitle = {Proceedings of the ACM SIGOPS 22nd Symposium on Operating Systems Principles},
  pages     = {207–220},
  numpages  = {14},
  location  = {Big Sky, Montana, USA},
  series    = {SOSP '09}
}

@inproceedings{Landgraf:2024:RVM,
  author    = {Landgraf, Joshua and Giordano, Matthew and Yoon, Esther and Rossbach, Christopher J.},
  title     = {Reconfigurable Virtual Memory for FPGA-Driven I/O},
  year      = {2023},
  isbn      = {9781450399180},
  publisher = {Association for Computing Machinery},
  address   = {New York, NY, USA},
  url       = {https://doi.org/10.1145/3582016.3582048},
  doi       = {10.1145/3582016.3582048},
  booktitle = {Proceedings of the 28th ACM International Conference on Architectural Support for Programming Languages and Operating Systems, Volume 3},
  pages     = {556–571},
  numpages  = {16},
  location  = {Vancouver, BC, Canada},
  series    = {ASPLOS 2023}
}

@inproceedings{Li:2021:FVMP,
  author    = {Shih-Wei Li and Xupeng Li and Ronghui Gu and Jason Nieh and John Zhuang Hui},
  title     = {Formally Verified Memory Protection for a Commodity Multiprocessor Hypervisor},
  booktitle = {30th USENIX Security Symposium (USENIX Security 21)},
  year      = {2021},
  isbn      = {978-1-939133-24-3},
  pages     = {3953--3970},
  url       = {https://www.usenix.org/conference/usenixsecurity21/presentation/li-shih-wei},
  publisher = {USENIX Association},
  month     = aug
}

@inproceedings{Li:2022:DVAC,
  author    = {Xupeng Li and Xuheng Li and Christoffer Dall and Ronghui Gu and Jason Nieh and Yousuf Sait and Gareth Stockwell},
  title     = {Design and Verification of the Arm Confidential Compute Architecture},
  booktitle = {16th USENIX Symposium on Operating Systems Design and Implementation (OSDI 22)},
  year      = {2022},
  isbn      = {978-1-939133-28-1},
  address   = {Carlsbad, CA},
  pages     = {465--484},
  url       = {https://www.usenix.org/conference/osdi22/presentation/li},
  publisher = {USENIX Association},
  month     = jul
}

@manual{linux:pvops,
  author       = {{The kernel development community}},
  title        = {{Paravirt\_ops}},
  organization = {{The Linux Kernel Organization}},
  year         = {2024},
  url          = {https://www.kernel.org/doc/html/latest/virt/paravirt_ops.html}
}

@manual{mackerel,
  author = {Roscoe, Timothy and {Barrelfish Projec}},
  title  = {{Mackerel User Guide -- Barrelfish Technical Note 2}}
}

@inproceedings{Mahfud:2023:SLR,
  author    = {Mahfud, Ahmad Zainudin and Sabila, Muhamad Tegar and Wibowo, Nugroho Adi and Rhamdhan, Agria and Priambodo, Dimas Febriyan},
  booktitle = {2023 3rd International Conference on Electronic and Electrical Engineering and Intelligent System (ICE3IS)},
  title     = {A Systematic Literature Review on Operating System Security: Distribution and Issues},
  year      = {2023},
  volume    = {},
  number    = {},
  pages     = {70-75},
  doi       = {10.1109/ICE3IS59323.2023.10335475}
}

@inproceedings{Markettos:2019:TEV,
  title     = {{Thunderclap: Exploring Vulnerabilities in Operating System IOMMU
               Protection via DMA from Untrustworthy Peripherals}},
  author    = {Markettos, A Theodore and Rothwell, Colin and Gutstein, Brett F and
               Pearce, Allison and Neumann, Peter G and Moore, Simon W and Watson, Robert NM},
  booktitle = {NDSS},
  year      = {2019}
}

@inproceedings{Markuze:2016:TIP,
  author    = {Markuze, Alex and Morrison, Adam and Tsafrir, Dan},
  title     = {{True IOMMU Protection from DMA Attacks: When Copy is Faster Than
               Zero Copy}},
  booktitle = {Proceedings of the Twenty-First International Conference on
               Architectural Support for Programming Languages and Operating Systems},
  series    = {ASPLOS '16},
  year      = {2016},
  isbn      = {978-1-4503-4091-5},
  location  = {Atlanta, Georgia, USA},
  pages     = {249--262},
  numpages  = {14},
  url       = {http://doi.acm.org/10.1145/2872362.2872379},
  doi       = {10.1145/2872362.2872379},
  acmid     = {2872379},
  publisher = {ACM},
  address   = {New York, NY, USA}
}

@inproceedings{McKee:2022:HAKC,
  author    = {McKee, Derrick and Giannaris, Yianni and Ortega, Carolina and Shrobe, Howard and Payer, Mathias and Okhravi, Hamed and Burow, Nathan},
  booktitle = {29th Annual Network and Distributed System Security Symposium},
  publisher = {The Internet Society},
  series    = {NDSS'22},
  year      = {2022},
  month     = {01},
  location  = {San Diego, California, USA},
  title     = {Preventing Kernel Hacks with HAKCs},
  doi       = {10.14722/ndss.2022.24026}
}

@manual{mips,
  title           = {{IDT79R4600 TM and IDT79R4700 TM
                     RISC Processor Hardware User's Manual}},
  optkey          = {key},
  author          = {{Integrated Device Technology, Inc.}},
  optorganization = {organization},
  optaddress      = {address},
  edition         = {Revision 2.0},
  month           = {April},
  year            = {1995},
  url             = {https://docslib.org/doc/718203/idt79r4600-and-idt79r4700-risc-processor-hardware-users-manual},
  optnote         = {note},
  optannote       = {annote}
}

@inproceedings{Morgan:2016:BIP,
  author    = {Morgan, Benot
               and Alata, Eric
               and Nicomette, Vincent
               and Kaaniche, Mohamed},
  booktitle = {2016 Seventh Latin-American Symposium on Dependable Computing
               (LADC)},
  title     = {{Bypassing IOMMU Protection against I/O Attacks}},
  year      = {2016},
  volume    = {},
  number    = {},
  pages     = {145-150},
  doi       = {10.1109/LADC.2016.31},
  issn      = {},
  month     = {Oct}
}

@article{Morgan:2018:IPIO,
  author  = {Morgan, Benot
             and Alata, Eric
             and Nicomette, Vincent
             and Kaaniche, Mohamed},
  title   = {{IOMMU Protection Against I/O Attacks: A Vulnerability and a Proof of
             Concept}},
  journal = {Journal of the Brazilian Computer Society},
  year    = {2018},
  month   = {Jan},
  day     = {09},
  volume  = {24},
  number  = {1},
  pages   = {2},
  issn    = {1678-4804},
  doi     = {10.1186/s13173-017-0066-7},
  url     = {https://doi.org/10.1186/s13173-017-0066-7}
}

@inproceedings{Nelson:2017:Hyperkernel,
  author    = {Nelson, Luke and Sigurbjarnarson, Helgi and Zhang, Kaiyuan and Johnson, Dylan and Bornholt, James and Torlak, Emina and Wang, Xi},
  title     = {Hyperkernel: Push-Button Verification of an OS Kernel},
  year      = {2017},
  isbn      = {9781450350853},
  publisher = {Association for Computing Machinery},
  address   = {New York, NY, USA},
  url       = {https://doi.org/10.1145/3132747.3132748},
  doi       = {10.1145/3132747.3132748},
  booktitle = {Proceedings of the 26th Symposium on Operating Systems Principles},
  pages     = {252–269},
  numpages  = {18},
  location  = {Shanghai, China},
  series    = {SOSP '17}
}

@inproceedings{Proskurin:2020:XMP,
  author    = {Proskurin, Sergej and Momeu, Marius and Ghavamnia, Seyedhamed and Kemerlis, Vasileios P. and Polychronakis, Michalis},
  booktitle = {2020 IEEE Symposium on Security and Privacy (SP)},
  title     = {xMP: Selective Memory Protection for Kernel and User Space},
  year      = {2020},
  volume    = {},
  number    = {},
  pages     = {563-577},
  doi       = {10.1109/SP40000.2020.00041}
}

@inproceedings{Rashid:1987:MIVA,
  author    = {Rashid, Richard and Tevanian, Avadis and Young, Michael and Golub, David and Baron, Robert and Black, David and Bolosky, William and Chew, Jonathan},
  title     = {Machine-independent virtual memory management for paged uniprocessor and multiprocessor architectures},
  year      = {1987},
  isbn      = {0818608056},
  publisher = {Association for Computing Machinery},
  address   = {New York, NY, USA},
  url       = {https://doi.org/10.1145/36206.36181},
  doi       = {10.1145/36206.36181},
  booktitle = {Proceedings of the Second International Conference on Architectual Support for Programming Languages and Operating Systems},
  pages     = {31–39},
  numpages  = {9},
  location  = {Palo Alto, California, USA},
  series    = {ASPLOS II}
}

@misc{Reid:2016:AASL,
  author       = {Alastair Reid},
  title        = {ARM's Architecture Specification Language},
  howpublished = {Online. \url{https://alastairreid.github.io/specification_languages/}},
  year         = {2016}
}

@inproceedings{Reid:2016:TSAS,
  author    = {Reid, Alastair},
  title     = {Trustworthy Specifications of ARM v8-A and v8-M System Level Architecture},
  year      = {2016},
  isbn      = {9780983567868},
  publisher = {FMCAD Inc},
  address   = {Austin, Texas},
  booktitle = {Proceedings of the 16th Conference on Formal Methods in Computer-Aided Design},
  pages     = {161–168},
  numpages  = {8},
  location  = {Mountain View, California},
  series    = {FMCAD '16}
}

@misc{Reid:2020:UAAI,
  author       = {Alastair Reid},
  title        = {Using ASLi with Arm's v8.6-A ISA specification},
  howpublished = {Online. \url{https://alastairreid.github.io/using-asli/}},
  year         = {2020}
}

@inproceedings{Ryzhyk:2009:ADD,
  author    = {Ryzhyk, Leonid and Chubb, Peter and Kuz, Ihor and Le Sueur, Etienne and Heiser, Gernot},
  title     = {Automatic Device Driver Synthesis with Termite},
  year      = {2009},
  isbn      = {9781605587523},
  publisher = {Association for Computing Machinery},
  address   = {New York, NY, USA},
  url       = {https://doi.org/10.1145/1629575.1629583},
  doi       = {10.1145/1629575.1629583},
  booktitle = {Proceedings of the ACM SIGOPS 22nd Symposium on Operating Systems Principles},
  pages     = {73–86},
  numpages  = {14},
  location  = {Big Sky, Montana, USA},
  series    = {SOSP '09}
}

@inproceedings{Ryzhyk:2014:UDD,
  author    = {Ryzhyk, Leonid and Walker, Adam and Keys, John and Legg, Alexander and Raghunath, Arun and Stumm, Michael and Vij, Mona},
  title     = {User-Guided Device Driver Synthesis},
  year      = {2014},
  isbn      = {9781931971164},
  publisher = {USENIX Association},
  address   = {USA},
  booktitle = {Proceedings of the 11th USENIX Conference on Operating Systems Design and Implementation},
  pages     = {661–676},
  numpages  = {16},
  location  = {Broomfield, CO},
  series    = {OSDI'14}
}

@inproceedings{Seshadri:2007:SecVisor,
  author    = {Seshadri, Arvind and Luk, Mark and Qu, Ning and Perrig, Adrian},
  title     = {SecVisor: a tiny hypervisor to provide lifetime kernel code integrity for commodity OSes},
  year      = {2007},
  isbn      = {9781595935915},
  publisher = {Association for Computing Machinery},
  address   = {New York, NY, USA},
  url       = {https://doi.org/10.1145/1294261.1294294},
  doi       = {10.1145/1294261.1294294},
  booktitle = {Proceedings of Twenty-First ACM SIGOPS Symposium on Operating Systems Principles},
  pages     = {335–350},
  numpages  = {16},
  location  = {Stevenson, Washington, USA},
  series    = {SOSP '07}
}

@inproceedings{Shi:2017:DeconstructingXen,
  title     = {Deconstructing Xen},
  author    = {Shi, Lei and Wu, Yuming and Xia, Yubin and Dautenhahn, Nathan and Chen, Haibo and Zang, Binyu and Guan, Haibing and Li, Jinming},
  booktitle = {Proceedings 2017 Network and Distributed System Security Symposium},
  year      = {2017},
  publisher = {Internet Society},
  series    = {NDSS'17},
  doi       = {10.14722/ndss.2017.23455}
}

@manual{smtlib2,
  title  = {The SMT-LIB Standard},
  author = {Clark Barrett and Pascal Fontaine and Cesare Tinelli},
  note   = {Version 2.6},
  year   = {2021},
  url    = {https://smt-lib.org/papers/smt-lib-reference-v2.6-r2021-05-12.pdf}
}

@inproceedings{Song::2016:HDFI,
  author    = {Song, Chengyu and Moon, Hyungon and Alam, Monjur and Yun, Insu and Lee, Byoungyoung and Kim, Taesoo and Lee, Wenke and Paek, Yunheung},
  booktitle = {2016 IEEE Symposium on Security and Privacy},
  title     = {HDFI: Hardware-Assisted Data-Flow Isolation},
  series    = {SP'16},
  year      = {2016},
  volume    = {},
  number    = {},
  pages     = {1-17},
  doi       = {10.1109/SP.2016.9}
}

@inproceedings{Song:2016:EKSI,
  author    = {Chengyu Song and
               Byoungyoung Lee and
               Kangjie Lu and
               William Harris and
               Taesoo Kim and
               Wenke Lee},
  title     = {Enforcing Kernel Security Invariants with Data Flow Integrity},
  booktitle = {23rd Annual Network and Distributed System Security Symposium},
  publisher = {The Internet Society},
  series    = {NDSS'16},
  location  = {San Diego, California, USA},
  year      = {2016},
  doi       = {10.14722/ndss.2016.23218}
}

@inproceedings{Wang:2015:SecPod,
  author    = {Wang, Xiaoguang and Chen, Yue and Wang, Zhi and Qi, Yong and Zhou, Yajin},
  title     = {SecPod: a framework for virtualization-based security systems},
  year      = {2015},
  isbn      = {9781931971225},
  publisher = {USENIX Association},
  address   = {USA},
  booktitle = {Proceedings of the 2015 USENIX Conference on Usenix Annual Technical Conference},
  pages     = {347–360},
  numpages  = {14},
  location  = {Santa Clara, CA},
  series    = {USENIX ATC '15}
}

@inproceedings{Wang:2023:Ghostwriter,
  author    = {Wang, Bingyao and Noorafshan, Sepehr and Achermann, Reto and Seltzer, Margo},
  title     = {Synthesizing Device Drivers with Ghost Writer},
  year      = {2023},
  isbn      = {9798400704048},
  publisher = {Association for Computing Machinery},
  address   = {New York, NY, USA},
  url       = {https://doi.org/10.1145/3623759.3624545},
  doi       = {10.1145/3623759.3624545},
  booktitle = {Proceedings of the 12th Workshop on Programming Languages and Operating Systems},
  pages     = {10–17},
  numpages  = {8},
  location  = {Koblenz, Germany},
  series    = {PLOS '23}
}

@inproceedings{Watson:2015:CHERI,
  author    = {Watson, Robert N. M. and Woodruff, Jonathan and Neumann, Peter G. and Moore, Simon W. and Anderson, Jonathan and Chisnall, David and Dave, Nirav and Davis, Brooks and Gudka, Khilan and Laurie, Ben and Murdoch, Steven J. and Norton, Robert and Roe, Michael and Son, Stacey and Vadera, Munraj},
  title     = {CHERI: A Hybrid Capability-System Architecture for Scalable Software Compartmentalization},
  year      = {2015},
  isbn      = {9781467369497},
  publisher = {IEEE Computer Society},
  address   = {USA},
  url       = {https://doi.org/10.1109/SP.2015.9},
  doi       = {10.1109/SP.2015.9},
  booktitle = {Proceedings of the 2015 IEEE Symposium on Security and Privacy},
  pages     = {20–37},
  numpages  = {18},
  series    = {SP '15}
}

@inproceedings{Witchel:2005:Mondrix,
  author    = {Witchel, Emmett and Rhee, Junghwan and Asanovi\'{c}, Krste},
  title     = {Mondrix: memory isolation for linux using mondriaan memory protection},
  year      = {2005},
  isbn      = {1595930795},
  publisher = {Association for Computing Machinery},
  address   = {New York, NY, USA},
  url       = {https://doi.org/10.1145/1095810.1095814},
  doi       = {10.1145/1095810.1095814},
  booktitle = {Proceedings of the Twentieth ACM Symposium on Operating Systems Principles},
  pages     = {31–44},
  numpages  = {14},
  location  = {Brighton, United Kingdom},
  series    = {SOSP '05}
}

@misc{arm:mte,
  author = {{Arm Ltd.}},
  howpublished = {White Paper},
  title = {Armv8.5-A -- Memory Tagging Extension},
  year = {2019}
}

@misc{arm:pac,
  author = {{Arm Ltd.}},
  howpublished = {Online. \url{https://developer.arm.com/documentation/109576/0100/Pointer-Authentication-Code/Introduction-to-PAC}},
  title = {Introduction to PAC},
  year = {2016}
}

\end{document}